\documentclass[aps,prb,twocolumn,notitlepage,showpacs,floatfix,superscriptaddress,10pt,
groupedaddress,superscriptaddress]{revtex4-1}
\usepackage{times,amsmath,amsfonts,amssymb,mathrsfs,graphics,graphicx,color,comment,bm}
\usepackage[next]{inputenc}
\usepackage[dvips]{epsfig}
\usepackage[pdfstartview=FitH]{hyperref}
\usepackage{pdfsync}
\usepackage{appendix}
\usepackage{epstopdf}
\def\be{\begin{equation}}
  \def\ee{\end{equation}}

\def\bi{\begin{itemize}}
  \def\ei{\end{itemize}}
\def\bn{\begin{enumerate}}
  \def\en{\end{enumerate}}
\def\bea{\begin{eqnarray}}
  \def\eea{\end{eqnarray}}

\def\ba{\begin{array}}
  \def\ea{\end{array}}
\def\bd{\begin{displaymath}}
  \def\ed{\end{displaymath}}

\begin{document}

\title{Quantum phase transition as an interplay of Kitaev and Ising interactions
}

\author{A. Langari}
\affiliation{Department of Physics, Sharif University of Technology, P.O.Box
  11155-9161, Tehran, Iran}
\affiliation{Center of excellence in Complex Systems and Condensed Matter
  (CSCM), Sharif
  University of Technology, Tehran 1458889694, Iran}
\affiliation{Max-Planck-Institut f\"ur Physik komplexer Systeme, 01187 Dresden,
  Germany}
\email{langari@sharif.edu}
\homepage{http://sharif.edu/~langari/}
\author{A. Mohammad-Aghaei}
\affiliation{Department of Physics, Sharif University of Technology, P.O.Box
  11155-9161, Tehran, Iran}
\author{R. Haghshenas}
\affiliation{Department of Physics, Sharif University of Technology, P.O.Box
  11155-9161, Tehran, Iran}


%
%






\date{\today}
\begin{abstract}
We study the interplay between the Kitaev and Ising interactions on both
ladder and two dimensional lattices.
We show that the ground state of the Kitaev ladder is a symmetry-protected topological (SPT) phase, which is
protected by a $\mathbb{Z}_2 \times \mathbb{Z}_2$ symmetry. It is confirmed by the degeneracy of the entanglement spectrum and
non-trivial phase factors (inequivalent projective representations of the symmetries),
which are obtained within infinite matrix-product representation of numerical
density matrix renormalization group. We derive the effective theory to describe the topological phase transition
on both ladder and two-dimensional lattices, which is given by the transverse field Ising model with/without
next-nearest neighbor coupling based on the primary Ising configurations. The ladder has three phases, namely,
the Kitaev SPT, symmetry broken ferro/antiferromagnetic order and classical spin-liquid.
The non-zero quantum critical point and its corresponding central charge are provided by the effective theory,
which
are in full agreement with the numerical results, i.e., the divergence of entanglement entropy
at the critical point, change of
the entanglement spectrum degeneracy and a drop in the ground-state fidelity.
The central charge of the critical points are either c=1 or c=2, with the magnetization
and correlation exponents being 1/4 and 1/2, respectively.
The transition from the classical spin-liquid phase
of the frustrated Ising ladder to the Kitaev SPT phase is mediated by a floating phase, which
shows strong finite entanglement scaling. In the absence of frustration, the 2D lattice shows
a topological phase transition from the $\mathbb{Z}_2$ spin-liquid state to the long-range ordered Ising phase at finite
ratio of couplings, while in the presence of frustration, an order-by-disorder transition is induced
by the Kitaev term. The 2D classical spin-liquid phase is unstable against the
addition of Kitaev term toward an ordered phase before the transition to the $\mathbb{Z}_2$ spin-liquid state.

\end{abstract}


\maketitle

\section{Introduction\label{introduction}}
Topologically ordered quantum many-body systems have received a great deal of interest due to rich
insights emerged from their nature, namely, lack of any local order parameter to characterize
them \citep{Tsui:1982, Hansson:2004}, i.e. failure of symmetry breaking paradigm, exhibiting
long-range entanglement \cite{Chen:2010}, robustness against local perturbations \cite{Nayak:2008, Kitaev:2003},
non-trivial anyon statistics \cite{Wen:1990, Wen:2007} and so on.
Topological quantum codes including color codes
~\cite{PhysRevLett.97.180501,PhysRevB.75.075103,PhysRevA.76.012305,PhysRevA.78.062312}
have a universal feature characterized by topological entanglement entropy~\cite{Kitaev:2006,Levin:2006}
manifesting their topological nature.
Emergent fermions and anyons~\cite{PhysRevLett.100.057208,PhysRevLett.100.177204,PhysRevB.80.075111}
are typical quasi-particle excitations above a topological ground-state that influence
the finite temperature properties with non-trivial limiting features~\cite{PhysRevA.80.012321}
and bound states~\cite{PhysRevB.89.045411}.
The stability of topological ordered state against thermal, external magnetic field~\cite{PhysRevB.88.035118} and other
interactions~\cite{PhysRevB.78.125102} (like Ising~\cite{Karimipour:2013}) is an interesting issue,
which could lead to the phase transition from
a topological state. For instance, an in-plane magnetic field on the toric code leads to both first- and 
second-order quantum phase transition~\cite{PhysRevB.79.033109}, while a perpendicular
magnetic field gives a first-order phase transition at the self-dual point of the effective
quantum compass model~\cite{PhysRevB.80.081104}. The 2D color code shows similar behavior
in the presence of a magnetic field~\cite{PhysRevB.87.094413} and Ising interactions~\cite{PhysRevB.88.214411}.
The nature of such phase transition and its corresponding quantum critical properties are debating issues
inherited from the topological properties of the model.

Recently, many efforts, inspired by the concepts of
quantum information theory, have been made to provide a comprehensive understanding of topological
order \cite{Pollmann:2010, XieChen-2011-Classification, XieChen:2011}. So far, it is believed there exists
three different kinds of topological orders, namely: symmetry protected topological (SPT) order,
long range entangled states with topological order and symmetry enriched topological order. The picture in 1D is complete
and the symmetry-fractionalization mechanism competently characterize the SPT phases.
%
In higher dimensions, it is believed that
symmetry fractionalization, symmetry breaking and
long-range entanglement mechanisms are capable of characterizing the
aforementioned orders. However, to get a complete understanding, further studies are currently active and demanding.

Characterization of topological order relies on appropriate non-local order parameters. In 1D,
entanglement spectrum distinguishes SPT orders from trivial ones \cite{Pollmann:2010}, while
non-local order parameters based on inequivalent projective representations \cite{Projective-note}
identifies different SPT orders \cite{Pollmann:2012}. In 2D, even so, there is not a
unique and faithful tool to classify them but topological entanglement entropy \cite{Levin:2006} is
assumed as the most common tool to characterize intrinsic topological phases.
The case of symmetry enriched topological
orders is more complicated as both orders, i.e. SPT and intrinsic topological orders,
simultaneously exist---some proposed non-local measures might hopefully identify them \cite{Huang:2013, Zaletel:2013}.
Understanding a quantum phase transition (QPT) from a topological phase to a trivial phase requires less
effort than the classification of phases, since local-order parameters according to symmetry breaking mechanism
can identify the quantum phase transition.
Novel quantum phase transitions, which rarely have been studied,
happen when there are two distinct topological phases.

In this article, we consider the Kitaev toric code model accompanied by different Ising interactions,
namely, rhombic-Ising (RI), leg-Ising (LI) and rhombic-leg-Ising (RLI) interactions, on ladder
and two-dimensional square geometries.
The non-frustrated RI case of ladder and 2D square lattice have been 
studied recently in Ref.~\onlinecite{Karimipour:2013}.
Here, we consider all possible Ising interactions, which include the frustrated models for both ladder and 2D lattices.
The Kitaev toric code \cite{Kitaev:2003} is a well-known model showing
topological order, while Ising model with respect to the geometry of lattice and
the type of interactions can show up symmetry broken phases and topological spin-liquid phases \cite{Balents:2010}.
The later is due to frustration of anti-ferromagnetic (AF) interactions of Ising model (AF RLI interactions),
which leads to a rich phase diagram
\cite{Arizmendi:1991,Sen:1992,Rieger:1996,Derian:2006,Beccaria:2006,Beccaria:2007,Chandra:2007,Chandra:2007a,Nagy:2011}.

We show that the ground state of the Kitaev ladder is an SPT phase by introducing the responsible
$\mathbb{Z}_2 \times \mathbb{Z}_2$ symmetries, which is confirmed by numerical results on the corresponding
 non-trivial phase factor \cite{Pollmann:2012}---see appendix-\ref{symmetry}.
To investigate the competition between the Kitaev SPT phase and an
Ising phase we employ two general approaches: an effective theory,
which comes from an exact map of the original model
to an effective one, and the infinite system density-matrix-renormalization-group (iDMRG)
algorithm \cite{Schollwock:2011, Vidal:2006, Mcculloch:2008} based on infinite matrix product state (iMPS) representation.
On the ladder geometry, the Kitaev plus Ising interactions are mapped to decoupled chains of
nearest-neighbor (NN) or next-nearest-neighbor (NNN) transverse field Ising (TFI) model.
The effective theory and numerical iDMRG computations show a quantum phase transition
at finite non-zero coupling from the Kitaev SPT phase to the broken symmetry antiferro/ferro-magnetic phase
except in the case of AF RLI interactions.
The existence of quantum critical point (QCP) and its location is proved by numerical simulation that leads to
the divergence of the entanglement entropy, change in the degeneracy of the entanglement spectrum, drop in the
ground-state fidelity, change of phase factor, and non-zero magnetic order parameter.
We have also computed the corresponding central charges, which is in agreement with the
proposed effective theory, namely, $c=2$ for Kitaev-LI, $c=1$ for both Kitaev-RI and Kitaev-RLI QCPs.
The critical exponents of magnetization and correlation function are $\beta=1/4$
and $\eta=1/2$, correspondingly for all types of non-frustrated ladders.
Concerning the QCP, our result for Kitaev-RI interactions is in contrast to
Ref.~\onlinecite{Karimipour:2013}, which concludes zero Ising
coupling strength, while we observe a finite non-zero Ising coupling ($J_z=J_v$, see Fig.~\ref{VNE_R}).
Moreover, the whole study for
the LI and RLI cases are new investigations
of this manuscript that includes the frustrated case.
The case of AF RLI coupling makes a competition between the Kitaev SPT phase and
a classical spin-liquid one, which can be explained in terms of the frustrated NNN TFI effective theory.
Our numerical results for AF RLI case are in favor of the existence of a floating phase
---which has
algebraic decaying correlations---between the
classical spin-liquid and Kitaev SPT phases. This is in agreement with the phase diagram
proposed for the (effective) frustrated NNN TFI chain in
Refs.~\onlinecite{Chandra:2007,Chandra:2007a}.

For 2D square lattice, we follow the same strategy and map the original model to
an effective theory, which is given by (NN or NNN) 2D TFI model in terms of (effective) quasi-spins.
The effective theory is defined on the bi-partite square lattice of quasi-spins, where the two sub-lattices
are decoupled.
For all types of Ising interactions except the case of AF Ising interactions on all bonds, 
we get a QPT at finite non-zero couplings
from the intrinsic topological ($\mathbb{Z}_2$ spin-liquid) phase to
a long-range ordered of ferro or antiferromagnetic (N\'{e}el) type.
At the extreme limit of AF Ising coupling on all bonds, the presence of frustration leads to
extensive degenerate configurations, which would be destabilized
by order-by-disorder transition~\cite{J.Physique.41.1263,JETP.56.178,PhysRevLett.84.4457} due to fluctuations induced by Kitaev term.
At higher values of Kitaev term a transition to
the topological $\mathbb{Z}_2$ spin-liquid phase occurs.

%

The remainder of this paper is organized as follows. In Sec.~\ref{Model} we first briefly
review the Kitaev toric code on ladder and introduce the $\mathbb{Z}_2 \times \mathbb{Z}_2$ symmetry, which
protects the degeneracy of the entanglement spectrum.
We then define different types of Ising interactions in Sec.~\ref{Ladder} and derive the effective
theory for the Kitaev-Ising interplay. We present our numerical results in the same section.
We consider the 2D version of the interplay between Kitaev and Ising interaction in Sec.~\ref{Twodimention},
where an effective theory is introduced for all types of Ising couplings.
Finally  in Sec.~\ref{Summary}, we end up with a summary and discussion.
The article is accompanied by two appendices,
which describe the underlying numerical iDMRG (iMPS) approach.

\section{The SPT phase of Kitaev ladder}
\label{Model}
\begin{figure}[tb!]
  \begin{center}
    \includegraphics[width=0.9 \linewidth]{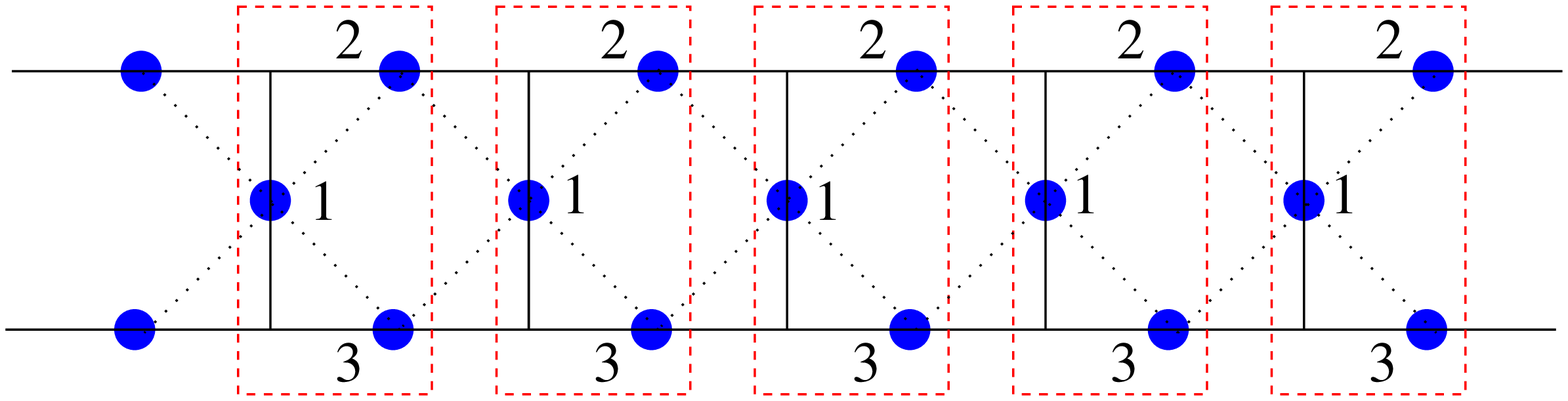}
    \caption{(color online) Two-leg Kitaev ladder, where the filled-blue circles show the position
      of real spins. The triangles show the vertex term and rhombuses represent the plaquette term
      in the Hamiltonian. The dashed-red rectangle shows the unit cell of the model.}
    \label{Kitaev-ladder}
  \end{center}
\end{figure}

The Kitaev ladder is defined on the ladder geometry as shown in Fig.~\ref{Kitaev-ladder}, where
the spins sit on the bonds of the two-leg ladder. The Kitaev Hamiltonian (${\cal H_K}$) is composed of
two terms, vertex ($A_v$) and plaquette  ($B_p$) interactions
\bea
{\cal H_K}= -J_v \sum_{\perp, \top} A_v -J_p \sum_{\square} B_p, \nonumber \\
A_v\equiv \prod_i \sigma_i^x, \hspace{2mm} i\in \perp \mbox{or } \top \;; \hspace{2mm}  J_v>0, \nonumber \\
B_p\equiv \prod_j \sigma_j^z, \hspace{2mm} j\in \Box \;; \hspace{2mm} J_p>0,
\label{HK}
\eea
where $\sigma^{\alpha}_j$ is the $\alpha$-component Pauli matrix at position $j$.
The model is exactly solvable \cite{Karimipour:2009}, which has two-fold topologically degenerate
ground states. Let $|\Omega\rangle \equiv \otimes_i |+ \rangle_i$, where $|+ \rangle_i$
is the eigenstate of $\sigma^x_i$, a ground state of Kitaev ladder is given by
\be
|\psi_{K}\rangle = \frac{1}{2^N}\prod_p (1+B_p) |\Omega\rangle,
\label{kgs}
\ee
where $N$ is the total number of rungs on the ladder.
And $|\psi'_{K}\rangle=W_z |\psi_{K}\rangle$ is the other ground state in which
$W_z=\prod_{\ell} \sigma^z_{\ell}$, where
$\ell$ runs only on one of the legs of ladder.
The ground state is understood as an equally weighted superposition of the states, which
are obtained by the operation of any homologically trivial loop of $\sigma^z$ operators on $|\Omega\rangle$.
The excited states can be constructed by the
operation of open strings of $\sigma^z$ operators on $|\Omega\rangle$.
A complete characterization of the spectrum shows that the excited states are at least two-fold
degenerate, which could be more except the highest energy level that has only
a double degeneracy \cite{Karimipour:2009}.
Moreover, the ground state entropy of the model is equal to $\ln(2)$.

The Kitaev ladder (${\cal H_K}$) has (i) two-fold degenerate ground state which can not be
distinguished by a local-order parameter of the Landau-Ginzburg symmetry breaking paradigm,
(ii) a finite-energy gap between the ground state and the first excited one and
(iii) anyonic excitations of integer magnetic and electric charges with Abelian statistics \cite{Karimipour:2009}.
Although the quasi-one dimensional Kitaev ladder does not bear topological characters
like Wilson loops and topological entanglement entropy
its ground state is classified to be an SPT phase.
We will show explicitly that the ground state of
Kitaev ladder is protected by a  $\mathbb{Z}_2\times \mathbb{Z}_2$  symmetry. For each unit cell of the two-leg ladder,
Fig.~\ref{Kitaev-ladder}, the following operator is defined,
\be
\Sigma^{abc}(j) = \sigma_1^{a}(j) \sigma_2^{b}(j) \sigma_3^{c}(j), \hspace{5mm} a, b, c=I, x, y, z,
\label{Sigma}
\ee
where $\sigma_{i}^I(j)$ is the identity operator at position $i$ of unit cell $j$.
It is straight forward to show that ${\cal H_K}$ is invariant under the operation of
the following two operators,
\be
{\cal X}=\prod_j \Sigma^{xxx}(j) \hspace{2mm} \mbox{and} \hspace{2mm} {\cal Z}=\prod_j \Sigma^{Izz}(j),
\label{XZ}
\ee
where $j$ runs over all unit cells. Moreover, $[{\cal X}, B_p]=0$, which states that the ground
state of Kitaev-ladder is invariant under ${\cal X}$. Similarly, it can be shown that the ground
state of Kitaev-ladder is invariant under ${\cal Z}$, since the product of ${\cal Z}$
by $\prod_p (1+B_p)$ is equivalent to the operation of $\prod_p (1+B_p)$. Hence, the
Kitaev-ladder ground state is invariant under the mutual symmetry operation ${\cal X} \times {\cal Z}$,
which defines the mentioned $\mathbb{Z}_2\times \mathbb{Z}_2$  symmetry.
The local symmetry operation $\Sigma^{Izz}, \Sigma^{xxx}$,
are two members of the group $G=\{\Sigma^{Izz}, \Sigma^{xxx}, -\Sigma^{xyy}, \Sigma^{III}\}$.
According to Ref.\onlinecite{Pollmann:2012},
we exploit this property and define an
order parameter ${\cal O}$, which can serve to detect, which projective
representation holds for the ground state in terms of its iMPS representation.
The order parameter ${\cal O}$ is defined by
\be
{\cal O} = \frac{1}{\chi} \text{Tr}\left(U_gU_{g'}U^{\dagger}_gU^{\dagger}_{g'}\right), \quad g\in G
\label{O}
\ee
where $U_g$ comes from the transformation of iMPS representation of the ground state ($\Gamma_{j}$)
under the symmetry ${\cal X} \times {\cal Z}$, i.e.  $\Gamma_{j}\rightarrow U^{\dagger}_{g}\Gamma_{j}U_{g}$,
where $g \in G$ and $\chi$ is the dimension of matrices in iMPS
(for details see appendix-\ref{symmetry}). If the ground state respects the symmetry,
${\cal O}=1$ for a trivial phase and ${\cal O}=-1$ for an SPT phase. Otherwise, when the ground state is not
invariant under symmetry we get ${\cal O}=0$, which shows the presence of symmetry breaking phenomenon.
${\cal O}$ is called ``phase factor order parameter". We find
numerically that the Kitaev phase
reveals ${\cal O}=-1$,
 (see Fig.~\ref{phases})
 which justifies that it is being protected by ${\cal X} \times {\cal Z}$
symmetries. Moreover, the entanglement spectrum is degenerate in the Kitaev phase
(see Fig.~\ref{ES_R}), which confirms that the ground state of Kitaev-ladder is an SPT phase.


\section{Kitaev-Ising ladder}
\label{Ladder}

\begin{figure}[tb!]
  \begin{center}
    \includegraphics[width=0.9 \linewidth]{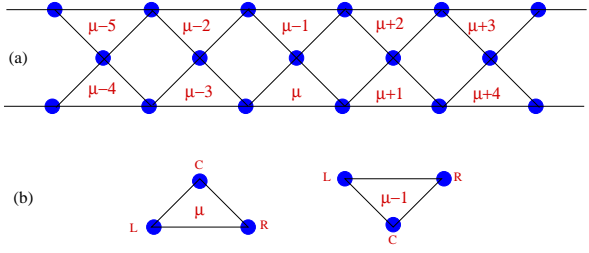}
    \caption{(color online) (a) The Kitaev ladder, where each vertex operator (triangle)
      is denoted as an effective spin ($\tau_{\mu}^z$) labeled by $\mu$. (b) The original
      spins are relabeled by the triangle index ($\mu$) and its position (L, R, C) on it.}
    \label{Effective-ladder}
  \end{center}
\end{figure}

The Ising term, which is composed of two-body interactions, competes with the SPT character
of the pure Kitaev ground state on the ladder. The Ising interaction $\sigma^z_i \sigma^z_j$, (which is defined on
the nearest neighbor spins of ladder) does not commute with the vertex terms ($A_v$) of ${\cal H_K}$
that establishes a competitions between a
symmetry-protected topological and a classical state. The classical state,
which is a result of strong Ising interaction could be realized in different forms according
to the pattern of Ising interactions. We classify three types of Ising interactions
on the two-legs ladder in Fig.~\ref{Kitaev-ladder}. (A) Rhombic-Ising interactions, where the
Ising terms are defined only between the nearest neighbor spins sitting on each rhombus.
The corresponding Hamiltonian (${\cal H_R}$) is defined in Eq.(\ref{H-rhombic}). (B) Leg-Ising
interactions, which is defined between nearest neighbor spins on the legs of ladder
and given by ${\cal H_L}$ in Eq.\ref{H-leg}. (C) Rhombic-leg Ising interactions that
is composed of nearest neighbor interaction between any pair of spins on the two-legs ladder,
which is being represented by the sum of two previous cases, i.e. ${\cal H_{RL}}={\cal H_R}+{\cal H_L}$.
We consider both ferromagnetic ($J_z >0$) and anti-ferromagnetic ($J_z <0$) coupling for the Ising terms.
The latter leads to
a rich structure of the ground state phase diagram as a result of frustration originated
from the anti-ferromagentic interactions on the bonds of triangles (see Fig.~\ref{Effective-ladder}).

To investigate the competition between Kitaev and Ising interactions, we introduce a transformation that gives the effective theory, which illustrates the
quantum phase transition of Kitaev model in the presence of Ising interactions.
The Hamiltonian is composed of three types of terms, i.e. the vertex ($A_v$), plaquette ($B_p$)
and Ising ($\sigma^z_i \sigma^z_j$) terms. The plaquette term  commutes with both vertex
and Ising ones, $[B_p, A_v]=0, [B_p, \sigma^z_i \sigma^z_j]=0$,
and consequently does not play any role
in the competition for quantum phase transition. However, the plaquette term adds a constant
term to the underlying Hamiltonian,
which is being fixed to its minimum value for the ground state properties, i.e. $B_p=+1$.

The building block of the effective theory is a triangle that is denoted by a vertex operator.
To visualize this picture, the ladder is labeled by its triangles corresponding to each
vertex operator in Fig.~\ref{Effective-ladder}-(a). In this representation, the spin of the original
lattice carries two indices, the label of triangle ($\mu$) and a label, which sticks to right (R), left (L)
or center (C) of a triangle, as can be seen in Fig.~\ref{Effective-ladder}-(b). A vertex operator is then
given by
\be
A_{\mu}=\sigma_{\mu, L}^x \sigma_{\mu, R}^x \sigma_{\mu, C}^x.
\ee
We consider the x-representation as the basis of our study.
In this representation, a vertex operator ($A_{\mu}$) has two values either $+1$
or $-1$, which is denoted by the associated quasi-spin ($\tau^z_{\mu}$), i.e.
\be
A_{\mu} \longrightarrow \tau^z_{\mu}.
\label{tauz}
\ee
It concludes that the effect of Kitaev Hamiltonian on the quasi-spin representation is
like a magnetic filed,
\be
J_v \sum_v A_v \longrightarrow J_v \sum_{\mu} \tau^z_{\mu}.
\label{kitaev-field}
\ee
The effect of a single $\sigma_{\mu, L}^z$ on a quasi-spin (a triangle) is to flip its state, which 
is denoted by $\tau^x_{\mu}$ in the quasi-spin representation,
\be
\sigma_{\mu, m}^z \longrightarrow \tau^x_{\mu} ,\;\;\;\; m=L, R, C.
\label{taux}
\ee
Depending on the geometry, where Ising interactions reside, we get three different effective models
leading to distinct critical points and universal behaviors.



\subsection{Rhombic-Ising interactions\label{RI-section}}

\begin{figure}[tb!]
  \begin{center}
    \includegraphics[width=0.9 \linewidth]{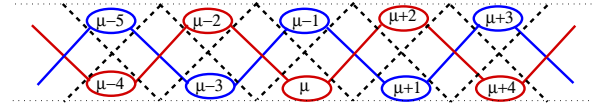}
    \caption{(color online) The effective model, which represents the Kitaev ladder in the
      presence of rhombic-Ising interactions that is composed of two decoupled Ising chains
      in a transverse field. An oval represents a quasi-spin ($\tau$).}
    \label{Rhombic-Ising}
  \end{center}
\end{figure}

The Rhombic-Ising terms are those two-body spin interactions, which act only along the edges of
Rhombic shapes in Fig.~\ref{Effective-ladder}-(a). According to the notation presented in
Fig.~\ref{Effective-ladder}-(b) the Ising Hamiltonian is given by
\be
{\cal H}_{\cal R}= -J_{z}\sum_{\mu} (\sigma^z_{\mu, C} \sigma^z_{\mu, R} + \sigma^z_{\mu, C} \sigma^z_{\mu, L}).
\label{H-rhombic}
\ee
It is important to note that the Rhombic-Ising terms do not change the state of a quasi-spin,
which shares an edge with the rhombus. For instance, $\sigma^z_{\mu, C} \sigma^z_{\mu, R}$ flips
two times the state of the triangle denoted by $\mu$, which leads to its original state
(see Fig.~\ref{Effective-ladder}-(b)).
However, the state of a triangle that only shares a single spin at its corner, is flipped. In other words,
the operation of $\sigma^z_{\mu, C} \sigma^z_{\mu, R}$ flips the state of quasi-spins (triangles)
denoted by $\mu-1$ and $\mu+1$.
Therefore, each Ising term, like $\sigma^z_{\mu, C} \sigma^z_{\mu, R}$ is represented by the product of
two x-component quasi-spin acting as $\tau^x_{\mu-1} \tau^x_{\mu+1}$. The Ising interactions on
the edges of a rhombus create effective interactions between the quasi-spins corresponding to the adjacent edges.
It leads to the effective interaction, $2 J_z \tau^x_{\mu-1} \tau^x_{\mu+1}$,
between two odd or two even quasi-spins, independently
(see Fig.~\ref{Rhombic-Ising}).
Thus, the effective Hamiltonian (${\cal H}_{eff}^{{\cal R}}$), which
describes the Kitaev Hamiltonian in the presence of Rhombic-Ising
interactions, ${\cal H_K}+{\cal H_R}$, is given by two decoupled
chain of transverse field Ising (TFI) model,
\be
{\cal H}_{eff}^{{\cal KR}}= -2 J_z \sum_{\mu}' \tau^x_{\mu} \tau^x_{\mu+2} - J_v \sum_{\mu}' \tau^z_{\mu},
\;\;\; \mu=\mbox{odd or even},
\label{H-eff-KR}
\ee
where $\sum_{\mu}'$ emphasizes the odd and even quasi-spins are decoupled.
Accordingly, a quantum phase transition takes place exactly at $2 J_z=J_v$, which
is known from the exact solution of spin-1/2 TFI chain. Our result is in contrast to
$J_z^c=0$ presented in Ref.\onlinecite{Karimipour:2013}.
In Ref.\onlinecite{Karimipour:2013} the Kitaev ladder with
Rhombic-Ising interaction is mapped to a spin-1/2 XY chain .using a non-local transformation.
Their mapping and the effective XY model are correct; however, the conclusion of zero Ising critical coupling
($J_z^c=0$) overlooks the true QCP. The exactly solvable spin-1/2 XY chain \cite{Lieb:1961} is defined by the Hamiltonian
$H_{XY}=J_x\sum_i ( s_i^x s_{i+1}^x + \gamma s_i^y s_{i+1}^y)$, where $\gamma\equiv J_y/J_x$.
For any non-zero value of $\gamma$ the XY chain is gapped except $\gamma=1$, where
the gap vanishes as $|\gamma - 1|$ at momentum $q=\pi/2$. The elementary excitations
at the (gapless) critical point are spinons \cite{Karbach:2008}.
Moreover, the 2nd derivative of ground state energy ($E_0$) diverges as
\be
\frac{d^2 E_0}{d \gamma^2} \sim |\gamma - 1|^{-3}  \hspace{5mm} \mbox{at} \hspace{5mm} q=\frac{\pi}{2},
\label{d2E0}
\ee
which justifies the quantum phase transition at the isotropic point, $\gamma=1$, that
corresponds to our result $J_z^c=J_v/2$.

\begin{figure}[tb!]
  \begin{center}
    \includegraphics[width=0.9 \linewidth]{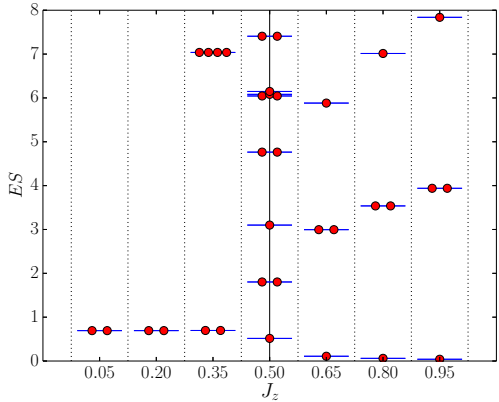}
    \caption{(color online) Entanglement spectrum (ES) versus $J_z$ for the ground state of ${\cal H_K}+{\cal H_R}$.
      The lowest level is doubly degenerate for $0\leq J_z < 0.5$, which is a signature of SPT
      character, while it is non-degenerate for $J_z > 0.5$ in the (trivial) ferromagnetic product state.
      At the quantum critical point $J_z=0.5$, the spectrum becomes dispersed over the entire range of
      eigenvalues, which is a signature of the critical point.}
    \label{ES_R}
  \end{center}
\end{figure}

To gain more insights on the structure of phases and the nature of quantum phase transition,
we obtain, by numerical calculations, the ground state of Kitaev ladder in the presence of
rhombic-Ising terms using an implementation of the iDMRG algorithm (see appendix-\ref{iDMRG}).
The code is based on iMPS representation, where
$\chi$ denotes the dimension of matrices in this formalism.
The entanglement spectrum of the ground state is defined in terms of the eigenvalues
of the reduced density matrix. Let $\rho$ be the ground-state reduced density matrix, which
is obtained by tracing over half of the ladder from the middle to either the left or right end of ladder,
\be
\rho = tr_{L/2}(|\psi_0\rangle \langle \psi_0|),
\label{rho}
\ee
where $|\psi_0\rangle$ is the ground state of Kitaev with Ising interactions. Let $\lambda_i$ be the eigenvalues of $\rho$, the entanglement
spectrum (ES) is defined by $\varepsilon_i= - \ln(\lambda_i)$. We have plotted the entanglement
spectrum versus $J_z$ of Kitaev ladder in the presence of rhombic-Ising interactions in Fig.~\ref{ES_R},
which exhibits a change of degeneracy at $J_z=0.5$.
We set $J_v=1$ as the scale of energy in all plots and results unless it appears explicitly.
The spectrum is doubly degenerate
for $0 \leq J_z < 0.5$, which is a clear signature for the SPT character of Kitaev phase, while
it is non-degenerate for $J_z>0.5$ in the (trivial) ferromagnetic product state. The change of degeneracy
of the entanglement spectrum at $J_z=0.5$ is an indication of quantum phase transition, which is accompanied
by a qualitative change in the ground state.

\begin{figure}[tb!]
  \begin{center}
    \includegraphics[width=0.99 \linewidth]{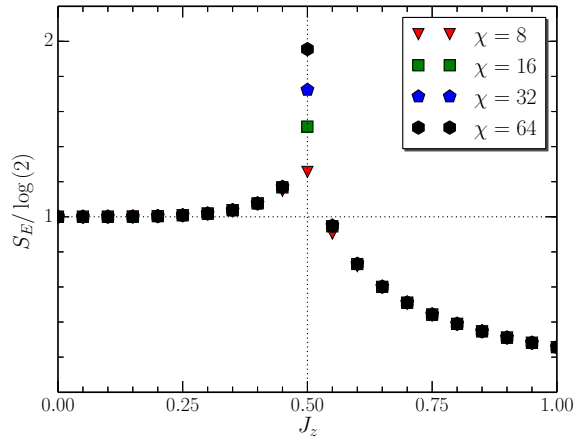}
    \caption{(color online) von Neumann entropy ($S_E$) versus $J_z$ for Kitaev plus rhombic-Ising interactions.
The divergent behavior of $S_E$ at $J_z=0.5$ is a clear signature of quantum phase transition. $S_E$ reaches $\ln(2)$ asymptotically
      for the pure Kitaev ladder ($J_z=0$).}
    \label{VNE_R}
  \end{center}
\end{figure}

The von Neumann (entanglement) entropy ($S_E$) is defined in terms of the eigenvalues of $\rho$,
\be
S_E= - \sum_{i} \lambda_i \ln(\lambda_i).
\label{VNE}
\ee
We have plotted $S_E$ versus $J_z$  in Fig.~\ref{VNE_R}
for different $\chi=8, 16, 32, 64$. The entropy shows a divergent behavior
only at $J_z=0.5$, which justifies the quantum phase transition. As shown in Fig.~\ref{VNE_R}, $S_E$
asymptotically reaches the value of $\ln(2)$ for the pure Kitaev ladder ($J_z=0$),
which is the
signature of its SPT character (the double degeneracy of ES), while it vanishes in the extreme Ising limit ($J_z \rightarrow \infty$)
representing a product state of up (or down) spins in a ferromagnetic state.

\begin{figure}[tb!]
  \begin{center}
    \includegraphics[width=0.9 \linewidth]{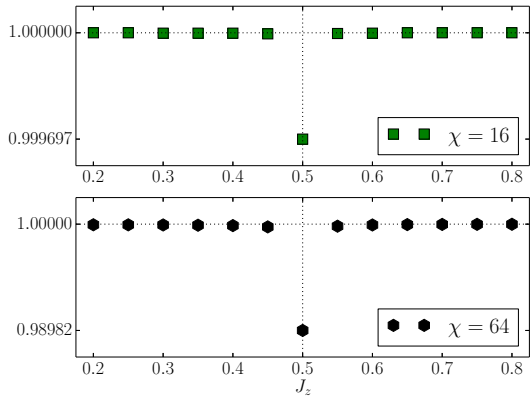}
    \caption{(color online) Ground-state fidelity versus $J_z$ for Kitaev-RI interactions.
      A clear drop at $J_z=0.5$ justifies the quantum phase transition.}
    \label{fid_R}
  \end{center}
\end{figure}

The ground-state fidelity (F) is a specific measure to investigate a quantum phase transition without an ad-hoc
assumption on the structure of ordering on any side of the transition point. The ground-state
fidelity is defined by
\be
F(J_z, J_z+\delta J_z)=\langle \psi_0(J_z) | \psi_0(J_z+\delta J_z) \rangle,
\label{fidelity}
\ee
where $\delta J_z$ is a very small amount of change in the coupling constant, which gives rise to
quantum phase transition. The ground-state fidelity is plotted in Fig.~\ref{fid_R} versus $J_z$
for two different values of $\chi=16, 64$  and $\delta J_z=0.01$. An obvious drop at $J_z=0.5$
confirms our observation of the quantum phase transition at $J_z=0.5$.

\begin{figure}[tb!]
  \begin{center}
    \includegraphics[width=0.9 \linewidth]{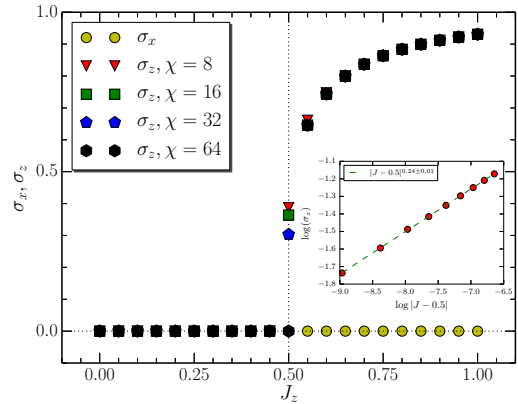}
    \caption{(color online) Magnetic order parameters versus $J_z$ for Kitaev plus RI interactions.
      The ferromagnetic order parameter, $\langle \sigma_z \rangle$ becomes nonzero for $J_z \geq 0.5$
justifying the quantum phase transition to the symmetry broken state. The inset shows the scaling
of $\langle \sigma_z \rangle \sim (J-0.5)^{(0.24\pm0.01)}$, where horizontal axis is in log-scale,
close to the critical point. }
    \label{S_R}
  \end{center}
\end{figure}

\begin{figure}[tb!]
  \begin{center}
    \includegraphics[width=0.9 \linewidth]{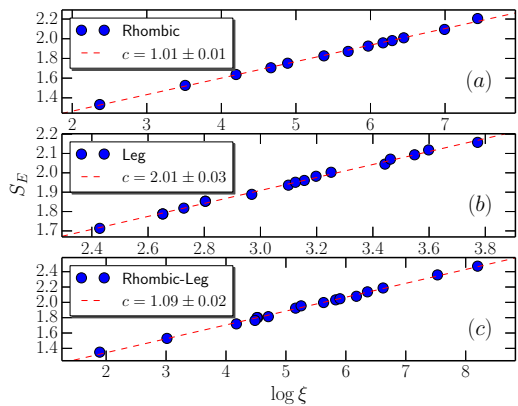}
    \caption{(color online) $S_E$ versus $\log(\xi)$, which gives the
central charge of the Kitaev ladder in addition to
 (a) RI, (b) LI and (c) RLI interactions according to the finite entanglement scaling .
(see appendix-\ref{iDMRG} and Ref.~\onlinecite{Kjall:2013,Calabrese:2004})}
    \label{charge}
  \end{center}
\end{figure}

We have also computed the ordinary magnetic order parameter on both sides of the quantum critical
point. We have plotted in Fig.~\ref{S_R} the magnetic order parameters in x and z directions, $\langle \sigma_x \rangle$
and $\langle \sigma_z \rangle$, respectively. $\langle \sigma_x \rangle$ is always zero for the whole
range of $J_z$, which shows no magnetic order in x-direction. However, $\langle \sigma_z \rangle$
becomes non-zero at $J_z=0.5$ indicating the magnetic order of the ferromagnetic state. Approaching the
quantum critical point from the ferromagnetic phase ($J_z > 0.5$) shows $\langle \sigma_z \rangle$
to vanish
like $(J-0.5)^{(0.24\pm0.01)}$ manifesting a second order phase transition
with exponent $\beta=0.24\pm0.01$ (the inset of Fig.~\ref{S_R}), in agreement with the effective theory
described in Eq.\ref{H-eff-KR}.
The effective theory, for Kitaev ladder with rhombic-Ising interactions,
is expressed in terms of two decoupled TFI chains, which predicts the central charge for the
corresponding QPT at $J_z=0.5$ is being twice the central charge of the TFI chain,
i.e. $c=2 \times 0.5=1$.
Similar argument shows that the magnetization exponent, which comes out of the effective
theory, is $\beta=1/4$ (which is discussed in Sec.\ref{Summary}).
We have numerically calculated the central charge at the
critical point $J_z=0.5$, which leads to $c=1.01\pm0.01$ as shown in
Fig.~\ref{charge}-(a).
The central charge is calculated
within finite-entanglement scaling introduced in Ref.\onlinecite{Kjall:2013}.
In this approach, the scaling of $S_E$ with the correlation length ($\xi$) would give a
fair approximation of the central charge,
(see appendix-\ref{iDMRG}). The numerical result confirms that the effective theory
truly captures the critical properties of the original model.

\begin{figure}[tb!]
  \begin{center}
    \includegraphics[width=0.9 \linewidth]{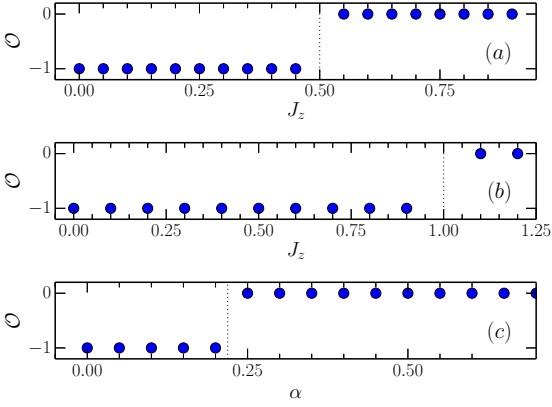}
    \caption{(color online) Phase factor order parameter for ferromagnetic Kitaev ladder in the
presence of (a) RI, (b) LI and (c) RLI interactions.}
    \label{phases}
  \end{center}
\end{figure}

We have plotted
the phase factor order parameter ${\cal O}$ versus $J_z$ in Fig.~\ref{phases}.
It shows that for small values of $J_z < J_z^c$, the model is in the SPT phase of Kitaev
ladder, which is justified by ${\cal O}=-1$. More specifically,  Fig.~\ref{phases}-(a) shows
that for $J_z < 0.5$ the model represents an SPT phase, while it shows a
symmetry broken trivial
phase (${\cal O}=0$) for  $J_z > 0.5$ via a quantum phase transition. The symmetry broken phase
does not respect the symmetry, which gives the largest eigenvalue of the transfer matrix to
be less than $1$, leading to ${\cal O}=0$.

Almost the whole discussion of the ferromagnetic rhombic-Ising
interaction is also valid for antiferromagnetic
Ising interaction. In other words, we can simply consider a mirror image of
all Figs. \ref{ES_R}, \ref{VNE_R},  \ref{fid_R}, \ref{S_R} and \ref{phases} with respect to $J_z=0$ to get
the antiferromagnetic regime. A better understanding can be achieved
by considering a $\pi$ rotation around x-axis
for the spins sitting on the rungs of ladder and $J_z \rightarrow - J_z$, which leaves the
whole Hamiltonian invariant.

\subsection{Leg-Ising interactions}

\begin{figure}[tb!]
  \begin{center}
    \includegraphics[width=0.9 \linewidth]{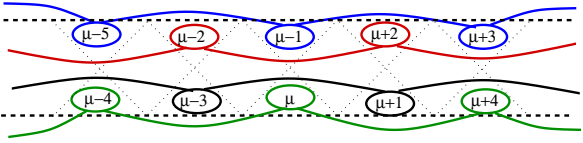}
    \caption{(color online) Schematic representation of the interactions in the effective Hamiltonian
      of Kitaev ladder in the presence of Leg-Ising interactions. The model is equivalent to four
      decoupled TFI chains.}
    \label{Leg-Ising}
  \end{center}
\end{figure}

The Ising terms may be considered only between the spins on the legs of ladder, without any
inter-leg interaction. Therefore, the Ising interactions would be between
spins labeled by L and R, namely: $\sigma^z_{\mu, R} \sigma^z_{\mu, L}$, which
is given by the following Hamiltonian,
\be
{\cal H}_{\cal L}= -J_{z}\sum_{\mu} \sigma^z_{\mu, L} \sigma^z_{\mu, R}.
\label{H-leg}
\ee
We explain the effect of $\sigma^z_{\mu, R} \sigma^z_{\mu, L}$ on quasi-spins.
The quasi-spin associated by $\mu$ is not changed by this Ising term as it flips two spins,
which leaves the product of spins on the triangle (quasi-spin) unchanged. However, the quasi-spins labeled
by $\mu-3$ and $\mu+1$ are being flipped (see
Fig.~\ref{Effective-ladder}), which
initiates  the following correspondence
in terms of quasi-spin operators,
\be
\sigma^z_{\mu, R} \sigma^z_{\mu, L} \longrightarrow \tau^x_{\mu-3} \tau^x_{\mu+1}.
\label{xx+4}
\ee
In accordance with Eq.\ref{xx+4}, the Ising interactions along the legs
are responsible for the interactions between the quasi-spins labeled mod(4, n), independently, where n=0, 1, 2, 3.
Therefore, the effective Hamiltonian is described by four decoupled TFI chains, namely
\be
{\cal H}_{eff}^{{\cal KL}}= - J_z \sum_{\mu}'' \tau^x_{\mu} \tau^x_{\mu+4} - J_v \sum_{\mu}'' \tau^z_{\mu},
\label{H-eff-KL}
\ee
where $\sum_{\mu}''$ indicates four decoupled chains as shown in Fig,\ref{Leg-Ising}.
The effective model shows a quantum phase transition at $J_z=J_v$.

\begin{figure}[tb!]
  \begin{center}
    \includegraphics[width=0.9 \linewidth]{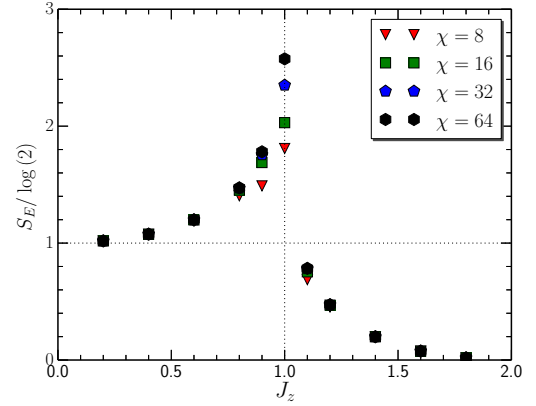}
    \caption{(color online) von Neumann entropy versus $J_z$ for the Kitaev ladder
      with leg-Ising interactions. The entropy diverges exactly at $J_z=1$ confirmed by the effective
      theory (${\cal H}_{eff}^{{\cal KL}}$).}
    \label{VNE_L}
  \end{center}
\end{figure}

The quantum phase transition at $J_z=1$ (for $J_v=1$) is justified by
von Neumann entropy versus $J_z$ plotted in Fig.~\ref{VNE_L}.
For the small Ising coupling ($J_z \rightarrow 0$)
$S_E$ is equal to $\ln(2)$ confirming the SPT phase of the pure Kitaev ladder.
The entropy rises up and becomes divergent at $J_z=1$, which is the signature of quantum phase
transition. Increasing the value of $J_z > 1$ leads to a ferromagnetic phase for the original
spins ordered in z-direction and mediated by the Ising interactions along the legs of ladder.
The factorized ferromagnetic state gives a zero value for $S_E$ as is shown in Fig.~\ref{VNE_L}
for the strong Ising coupling ($J_z \rightarrow \infty$).

Another indication of quantum phase transition is found in the structure of entanglement spectrum and
specially the degeneracy of levels. The degeneracy of the entanglement spectrum is even for $J_z <1$ while
it becomes odd for $J_z >1$ (not shown here). The type of spectrum is similar to Fig.~\ref{ES_R} except the
change of degeneracy, which occurs at $J_z=1$. The even degeneracy for $J_z <1$ is a signature of
the Kitaev SPT phase, which is verified by ${\cal O}=-1$ in Fig.~\ref{phases}-(b). The phase factor order parameter
(${\cal O}$) jumps to zero for $J_z >1$.

Our numerical results show that the quantum critical point between Kitaev SPT and ferromagnetic
leg-Ising phase is described by the central charge $c=2.01\pm0.03$ as shown in Fig.~\ref{charge}-(b).
This is in agreement with the effective theory obtained in Eq.\ref{H-eff-KL}, which shows
four decoupled TFI chains that give $c=4 \times 0.5=2$.

The phase diagram of the Kitaev ladder with antiferromagnetic leg-Ising interaction is
the mirror image of the ferromagentic phase diagram with respect to $J_z=0$. In fact, the
full Hamiltonian is invariant
under the transformation $J_z \rightarrow - J_z$ and
a $\pi$ rotation around x-axis on the even (or odd) spins on the legs of ladder.


\subsection{Rhombic-Leg-Ising interactions}

\begin{figure}[tb!]
  \begin{center}
    \includegraphics[width=0.9 \linewidth]{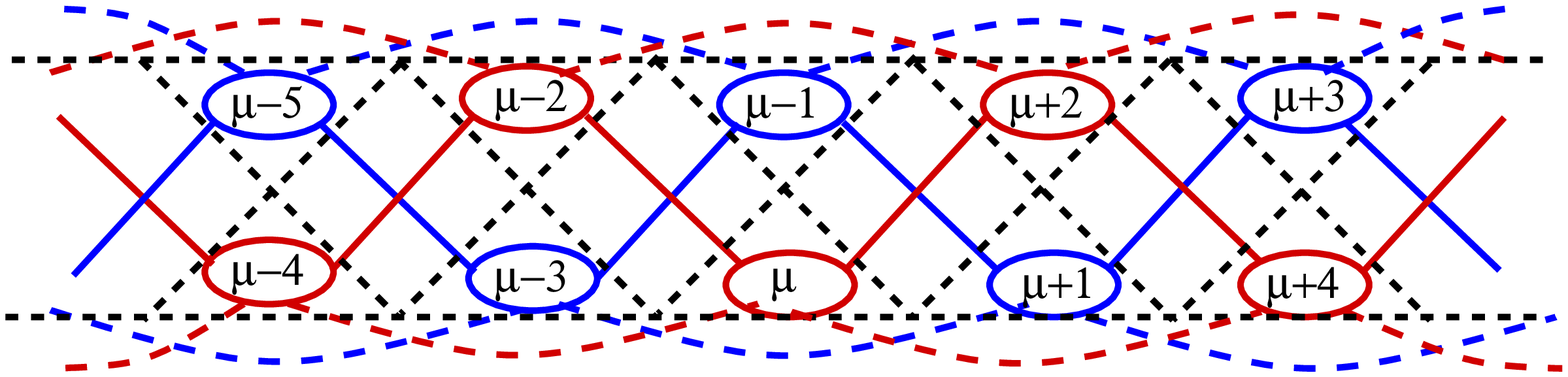}
    \caption{(color online) The effective interactions between the quasi-spins for the Kitaev ladder
      with both rhombic and leg Ising terms. The rhombic terms create two decoupled TFI chains, namely odd and even ones,
      where the leg terms add the next-nearest neighbor interactions on each chain, separately. Solid-red (blue) lines
      represent the nearest neighbor interactions, while dashed-red (blue) line show the next-nearest
      neighbor ones, on even (odd) quasi-spins.}
    \label{Rhombic-Leg-Ising}
  \end{center}
\end{figure}

We consider the Ising interactions along
the legs and the rhombic plaquettes that leads to more interesting phase diagram, where the full Hamiltonian is given by
\be
{\cal H_{KRL}}=(1-|\alpha|){\cal H_K} + \alpha ({\cal H_{R}} + {\cal H_{L}}), \hspace{2mm} |\alpha|\leq 1.
\label{KRL}
\ee
Here, we introduce $\alpha$ to sweep between the extreme limits of Kitaev interaction for $\alpha=0$
and Ising limit for $|\alpha|=1$. $\alpha >0$ corresponds to the ferromagnetic Ising interactions,
while $\alpha <0$ represents the anti-ferromagnetic ones.
The ground state phase diagram can be understood in terms of competition between the nearest and next-nearest
neighbors interactions, which comes out of the effective theory. 
The antiferromagnetic Ising interactions have specific features,
where frustration hinders simultaneous minimization of energy according to a
classical antiferromagnetic state. The effective theory
is simply obtained by incorporating the representation of rhombic-Ising and leg-Ising interactions in
the quasi-spin representations. The Ising terms on rhombus lead to Ising interaction between even (odd) quasi-spins, while
the leg terms establish interactions between quasi-spins of $\mu$ and $\mu+4$. Hence, the even and odd chain of
quasi-spins remain decoupled bearing the next-nearest neighbor interactions, which are the effect of leg-Ising
interactions. This can be seen in Fig.~ \ref{Rhombic-Leg-Ising}, where solid-red (blue) lines show
NN and dashed-red (blue) lines represent NNN interactions for the even (odd)
decoupled effective chains. It should be mentioned that the strength of NNN coupling is half of the
NN one. The effective Hamiltonian for the Kitaev ladder in the presence of rhombic-leg-Ising interactions
is given by
\bea
{\cal H}_{eff}^{{\cal KRL}}=&& - \alpha J_z \sum_{\mu}' (2 \tau^x_{\mu} \tau^x_{\mu+2} + \tau^x_{\mu} \tau^x_{\mu+4})
\nonumber \\
&&-(1-|\alpha|) J_v \sum_{\mu}' \tau^z_{\mu}, \hspace{3mm} \mu=\mbox{odd or even}.
\label{H-eff-KLR}
\eea
The presence of NNN interactions in the antiferromagnetic regime lead to the
interesting and exotic features in the model. Thus, we discuss the ferromagnetic and antiferromagnetic
cases in the following two subsections, separately.

\subsubsection{Ferromagnetic RL-Ising}

\begin{figure}[tb!]
  \begin{center}
    \includegraphics[width=0.9 \linewidth]{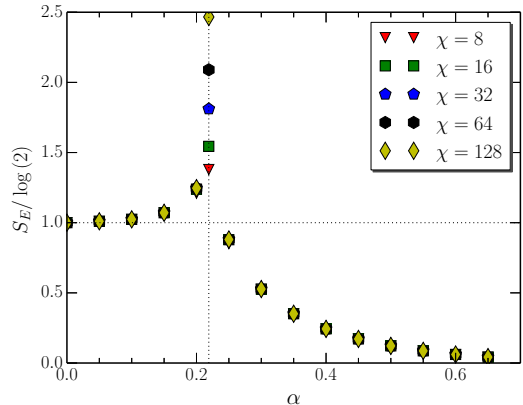}
    \caption{(color online) von Neumann entropy (scaled by $\ln(2)$) versus $\alpha$ for the Kitaev ladder in the presence of
      ferromagnetic rhombic-leg Ising interactions.}
    \label{VNE_RL_F}
  \end{center}
\end{figure}

Contrary to the RI and LI cases, the effective theory for the Kitaev ladder in the
presence of both leg and rhombic Ising terms does not have an exact solution due to the
NNN coupling in the TFI effective chain. Our numerical simulation of entropy, $S_E$ versus $\alpha$
is plotted in Fig.~\ref{VNE_RL_F}, which shows divergent behavior at the
critical point $\alpha_c^{{\cal KRL}}=0.219\pm 0.001$. This is equivalent to a phase
transition at $(J_z/J_v)=0.280\pm 0.001$ with a rescaling of the Kitaev and Ising couplings in Eq.~\ref{KRL}.
Here, the presence of Ising interactions on all bonds (legs and rhombuses) sustain the
ferromagnetic order to overcome the Kitaev SPT phase within smaller $J_z$ coupling
than the RI and LI cases.
The model represents the Kitaev SPT phase for $\alpha < \alpha_c^{{\cal KRL}}$, with finite
entanglement entropy $\ln(2)$, double degeneracy in the lowest entanglement spectrum,
no (local) magnetic order, and phase factor order parameter ${\cal O}=-1$ in Fig.~\ref{phases}-(c) .
A second order phase transition drives the model to the (trivial)
ferromagnetic phase for $\alpha > \alpha_c^{{\cal KRL}}$, which is presented by a factorized state of
up (or down) spins in z-direction.
The quantum critical point is described by the central charge $c=2 \times 0.5=1$ given by two decoupled NNN TFI chain and
justified by numerical simulation in Fig.~\ref{charge}-(c), which renders $c=1.09\pm0.02$.

\begin{figure}[tb!]
  \begin{center}
    \includegraphics[width=0.9 \linewidth]{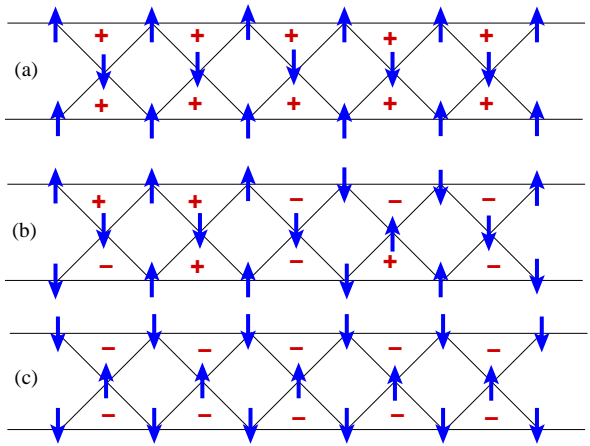}
    \caption{(color online) Some of ground-state configurations at the antiferromagnetic Ising limit ($J_v=0$) of
      the two-legs ladder. The $\uparrow$ and $\downarrow$ represent $\sigma^z$ spin orientation. The $\pm$
      shows the sum of $\sigma^z$ spins in each triangle. All configurations are classified as
      2-up-1-down or 2-down-1-up for each triangle. }
    \label{AF-Ising-limit}
  \end{center}
\end{figure}

\subsubsection{Antiferromagnetic RL-Ising}

The antiferromagnetic Ising interactions on both legs and rhombus bonds
create the basic building block of frustrated magnetic systems, i.e., triangles
with antiferromagnetic bonds (see Fig.~\ref{Effective-ladder}).
The antiferromagnetic Hamiltonian is defined by Eq.~\ref{KRL} with $-1 \leq \alpha \leq 0$.
Although the ground state of the model is two-fold degenerate at the Kitaev
limit ($\alpha=0$) it has exponentially degenerate ground state configurations
at the AF Ising limit ($\alpha=-1$).
To get more insight on the model
at the AF Ising limit, we
associate a magnetization ($m^z_{\mu}$) to each triangle,
which is simply the total magnetization in z-direction of a single triangle.
For the ground state, the antiferromagnetic nature of interactions enforce the spins
on each triangle
to be oriented as either 2-up-1-down or 2-down-1-up (see Fig.~\ref{AF-Ising-limit}),
which yields $m^z_{\mu}=\pm 1$. Therefore, the ground-state degeneracy at the
AF Ising limit is $2^{2N}$, where $2N$ is the number of triangles in the ladder
(assuming periodic boundary condition along legs).
The spins that sit on the legs of ladder are not constraint to a boundary
condition perpendicular to the legs of ladder, which leads to an intensive
degenerate configurations with total magnetization $M^z \in \{2N, 2N-1, \dots, -2N+1, -2N \}$,
\be
M^z= \sum_{\mu=1}^{2N} m^z_{\mu}.
\label{Mz}
\ee
Some configurations of the mentioned subspace are shown in Fig.~\ref{AF-Ising-limit}, where
the $\pm$ in each triangle represents $m^z_{\mu}$. A state with $M^z=2N$
is shown in Fig.~\ref{AF-Ising-limit}-(a),
where all triangles carry $m^z_{\mu}=+$, an intermediate
state with $M^z=0$ is presented in Fig.~\ref{AF-Ising-limit}-(b) and a state of
all  $m^z_{\mu}=-$
is given in Fig.~\ref{AF-Ising-limit}-(c). Accordingly, the model
does not show a magnetic long-range order out of a symmetry breaking, which is
called a {\it classical spin-liquid}.

\begin{figure}[tb!]
  \begin{center}
    \includegraphics[width=0.9 \linewidth]{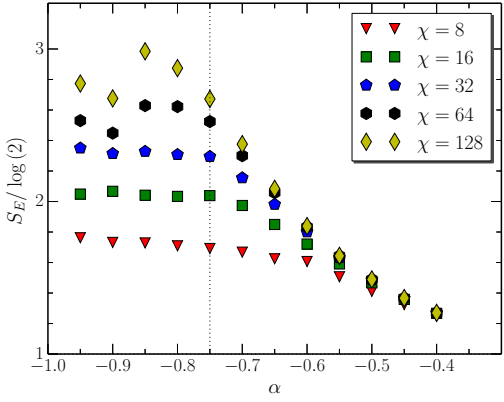}
    \caption{(color online) von Neumann entropy versus $\alpha$ for the Kitaev ladder in the presence of
      AF RL Ising interactions.}
    \label{VNE_RL_A}
  \end{center}
\end{figure}

\begin{figure}[tb!]
  \begin{center}
    \includegraphics[width=0.49 \linewidth]{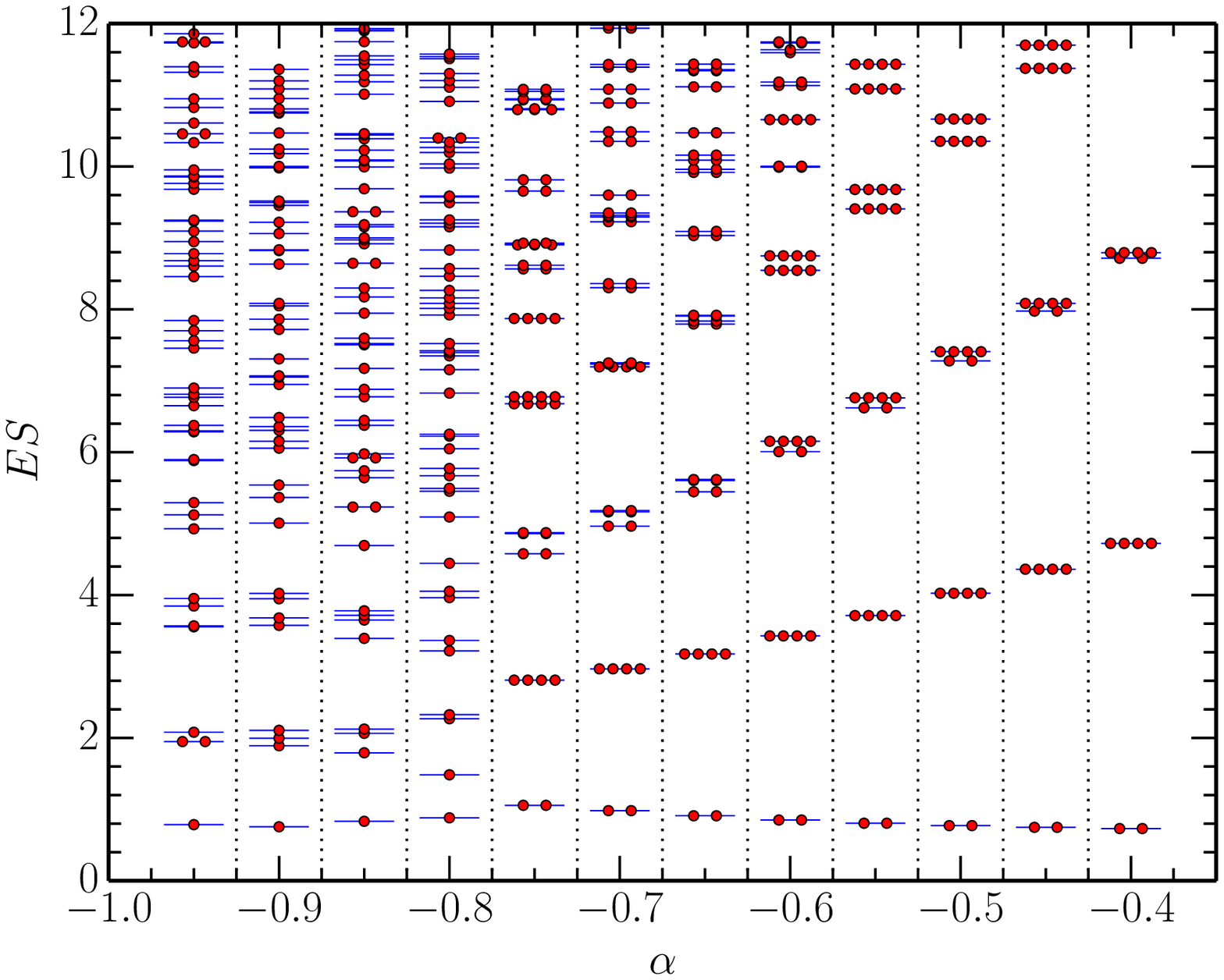}
    \includegraphics[width=0.49 \linewidth]{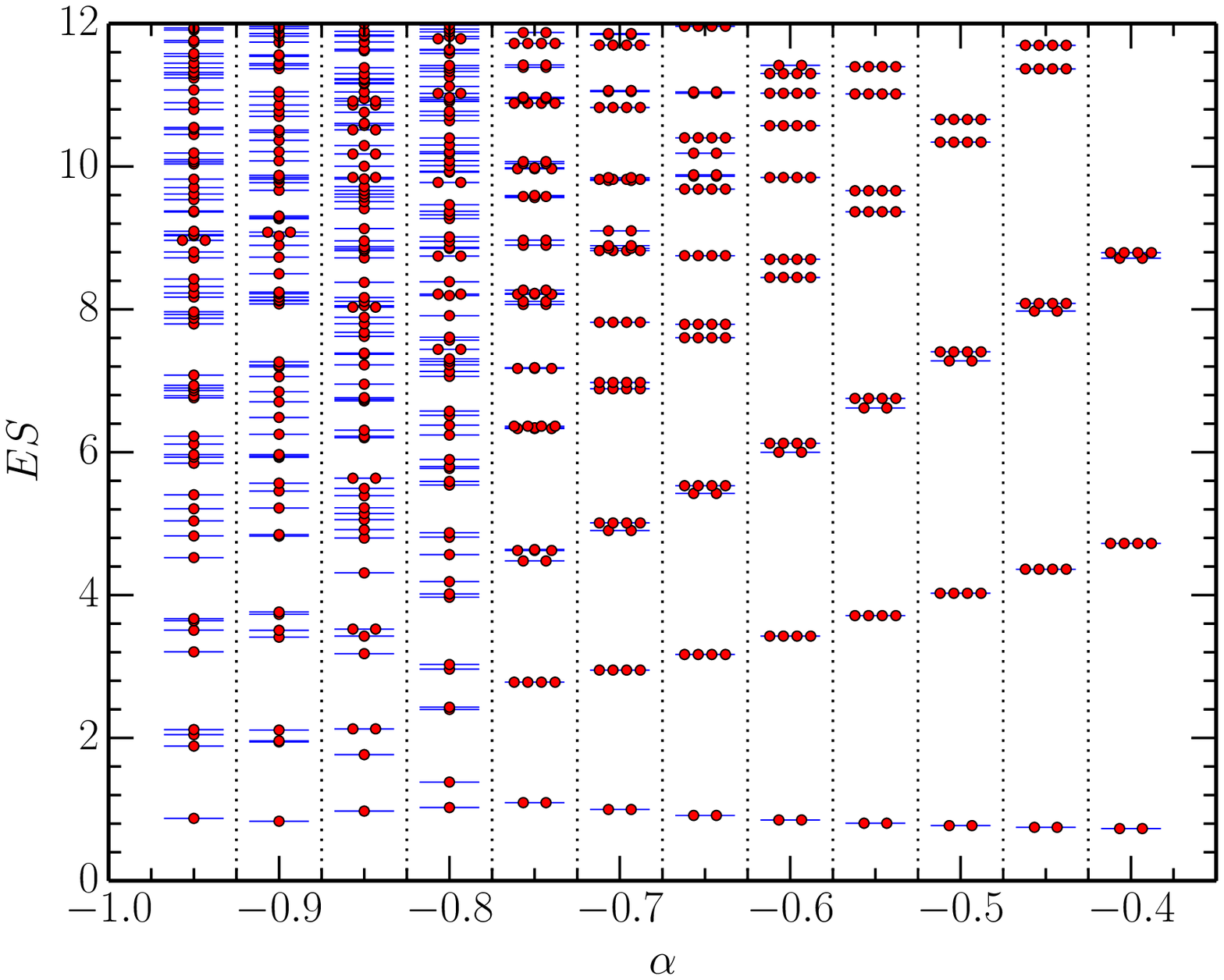}
    \caption{(color online) Entanglement spectrum versus $\alpha$ for the Kitaev ladder in the presence of
      AF RL Ising interactions. (Left): $\chi=64$, (right): $\chi=128$.}
    \label{ES_RL_A}
  \end{center}
\end{figure}

We have plotted $S_E$ versus $\alpha$ in Fig.~\ref{VNE_RL_A} for
Kitaev ladder in the presence of AF RL-Ising interactions. Our data shows
that $S_E$ reaches $\ln(2)$ for $\alpha \rightarrow 0$, which is the signature of
Kitaev SPT phase. The entanglement entropy shows finite entanglement scaling
for $-1 < \alpha < -0.7$, which is represented by $\chi=8, 16, 32, 64, 128$. Although
a bump is observed around $\alpha \simeq -0.8$ the whole set of data does not conclude to
a single divergent peak, rather showing a broad area of finite entanglement scaling.
It suggests a broad critical area, which starts at $\alpha=-1$ (the classical spin-liquid) toward
an intermediate region, $\alpha \simeq -0.75$, where the Kitaev SPT phase dominates.
This is confirmed by the structure of entanglement spectrum versus $\alpha$ presented in Fig.~\ref{ES_RL_A}.
We have shown in Sec.~\ref{Model} that the ground state of Kitaev ladder is an SPT phase, which
leads to even degeneracy of ES. Accordingly, the even degeneracy of ES is the signature of Kitaev SPT phase
for $\alpha_c < \alpha \leq 0$, where $\alpha_c=-0.75 \pm 0.05$. The computation for
higher values of $\chi$ ($>128$) is a massive time consumption for our model, where the unit
cell contains three spin-1/2. However, the results shown in Fig.~\ref{ES_RL_A} for $\chi=64$ (left),
$\chi=128$ (right) and other $\chi=16, 32$ (not shown here) convince us that the degeneracy of
Kitaev SPT is persistent for $\alpha_c < \alpha \leq 0$.
For $-1 < \alpha \leq \alpha_c$, the model shows finite entanglement scaling (Fig.~\ref{VNE_RL_A})
and dispersed ES (Fig.~\ref{ES_RL_A}), which resembles a critical
area with algebraic decay of correlation functions. This is consistent with the conclusion that can be derived
from the effective theory.

The effective Hamiltonian defined in Eq.\ref{H-eff-KLR} for Kitaev ladder
in the presence of AF RL-Ising interactions renders the frustrated NNN TFI chain in which
the NNN coupling is half of the NN one, being denoted by $\kappa=0.5$.
The effective theory at $\alpha=-1$ falls exactly on the critical point $\kappa=0.5$ at zero transverse field, which
separates the antiferromagnetic phase from the anti-phase of frustrated NNN TFI
\cite{Rieger:1996,Chandra:2007,Beccaria:2007,Nagy:2011}. It states that the classical spin-liquid
of the AF RL-Ising limit corresponds to the critical point of frustrated NNN TFI model at zero transverse field.
The onset of Kitaev term ($\alpha \neq -1$) adds quantum fluctuations to the model, which corresponds
to the effect of transverse field on the frustrated NNN TFI critical point. A recent study on the frustrated NNN TFI
chain \cite{Nagy:2011} confirms the existence of a tri-critical point at $\kappa=0.5$, where
a Kosterlitz-Thouless transition line and two second order transition line merge at $\kappa=0.5$ and zero field.
Thus, the effect of Kitaev term on the classical spin-liquid is similar to the effect of transverse field
on frustrated NNN TFI at $\kappa=0.5$ toward passing through the floating phase before reaching
a paramagnetic phase. The floating phase has algebraic decaying correlation functions, which could lead
to finite entanglement scaling and broad dispersion of ES.
Therefore, our results in Figs.~\ref{VNE_RL_A}-\ref{ES_RL_A} are in agreement with
the phase diagram proposed in Refs.~\onlinecite{Chandra:2007,Chandra:2007a} for NNN TFI (that
is usually denoted by ANNNI model in the literature).

\section{Two dimensional Kitaev-Ising model}
\label{Twodimention}
The extension of our model to two dimension would lead to the Kitaev toric code
Hamiltonian \cite{Kitaev:2003} in the presence of two-body Ising interactions.
The toric code is defined on the two-dimensional square lattice, where the spins
sit on the bonds of lattice (filled-black circles in Fig.~\ref{Effective-2D}-(a)).
The toric code Hamiltonian (${\cal H}_T$) is composed of two terms, vertex and plaquette ones, similar
to Eq.\ref{HK},
\be
{\cal H}_T= -J_v \sum_{+} A_v -J_p \sum_{\square} B_p,
\label{toric}
\ee
while both $A_v$ and $B_p$ are four-body interactions of $\sigma^x$ and
$\sigma^z$ operators around each vertex and plaquette, respectively.
The Ising interaction is composed of two-body interactions $- J_v \sigma^z_i \sigma^z_j$, where
$(i, j)$ represent the nearest neighbor spins (filled-black circles in Fig.~\ref{Effective-2D}-(a)).
Similar to the ladder cases, we
consider three types of Ising interactions, A: diagonal (D), where $(i, j)$ show the bonds
along diagonal directions (on the rhombic shapes of Fig.~\ref{Effective-2D}-(a)), B: horizontal-vertical (HV) that
defines the Ising bonds only on the horizontal and vertical links, C: full-Ising (FI), in which the
Ising interactions exist on all diagonal, horizontal and vertical bonds of the model.
The toric code in the presence of diagonal-Ising terms has been studied by Karimipour, et.al \cite{Karimipour:2013},
where a non-local transformation of bases map the model into a TFI model on a square lattice for two decoupled
sublattices. Moreover, a mean-field approximation predict a quantum phase transition at $J_v= 8J_z$.
Here, we implement the quasi-spin transformation, which creates the opportunity to study not only the diagonal-Ising
case, but also the other two cases, namely: HV and full-Ising models. It is very important to mention that
the full-Ising interaction introduces a frustrated magnetic system, which would give rise to exotic phases.

The quasi-spin representation for the two-dimensional model is defined on each vertex, where a vertex operator
$A_v=\sigma^x_1 \sigma^x_2 \sigma^x_3 \sigma^x_4$ results in either $+1$ or $-1$ in the $\sigma^x$-representation.
(The indices $1, 2, 3, 4$ represent the four spins sharing a vertex).
Thus, a quasi-spin is associated to this two-valued operator, which is called $T^z$,
\be
A_v \longrightarrow T^z.
\label{Omega}
\ee
As before, the plaquette term ($B_p$) commutes with both vertex and any types of Ising interaction.
Therefore, $B_p$ does not participate in the competition of a quantum phase transition and is being kept
to its minimum value to insure being in the low-energy sector.
Hence, the toric code Hamiltonian is represented by a magnetic field within the quasi-spin representation,
\be
{\cal H}_T= -J_v \sum_{+} A_v \longrightarrow -J_v \sum_{\nu} T^z_{\nu},
\label{effective-toric}
\ee
where $\nu$ runs over all vertices of the two-dimensional lattice. An Ising interaction
flips the value of a quasi-spin if it shares one spin with the vertex and do nothing if it shares two spins
with the vertex. Based on the type of Ising interaction we find different effective Hamiltonians
for ${\cal H}_{2D}={\cal H}_T+{\cal H}_{Ising}$, which
are given in the following subsections.

\subsection{Diagonal Ising interactions}

\begin{figure}[tb!]
  \begin{center}
    \includegraphics[width=0.9 \linewidth]{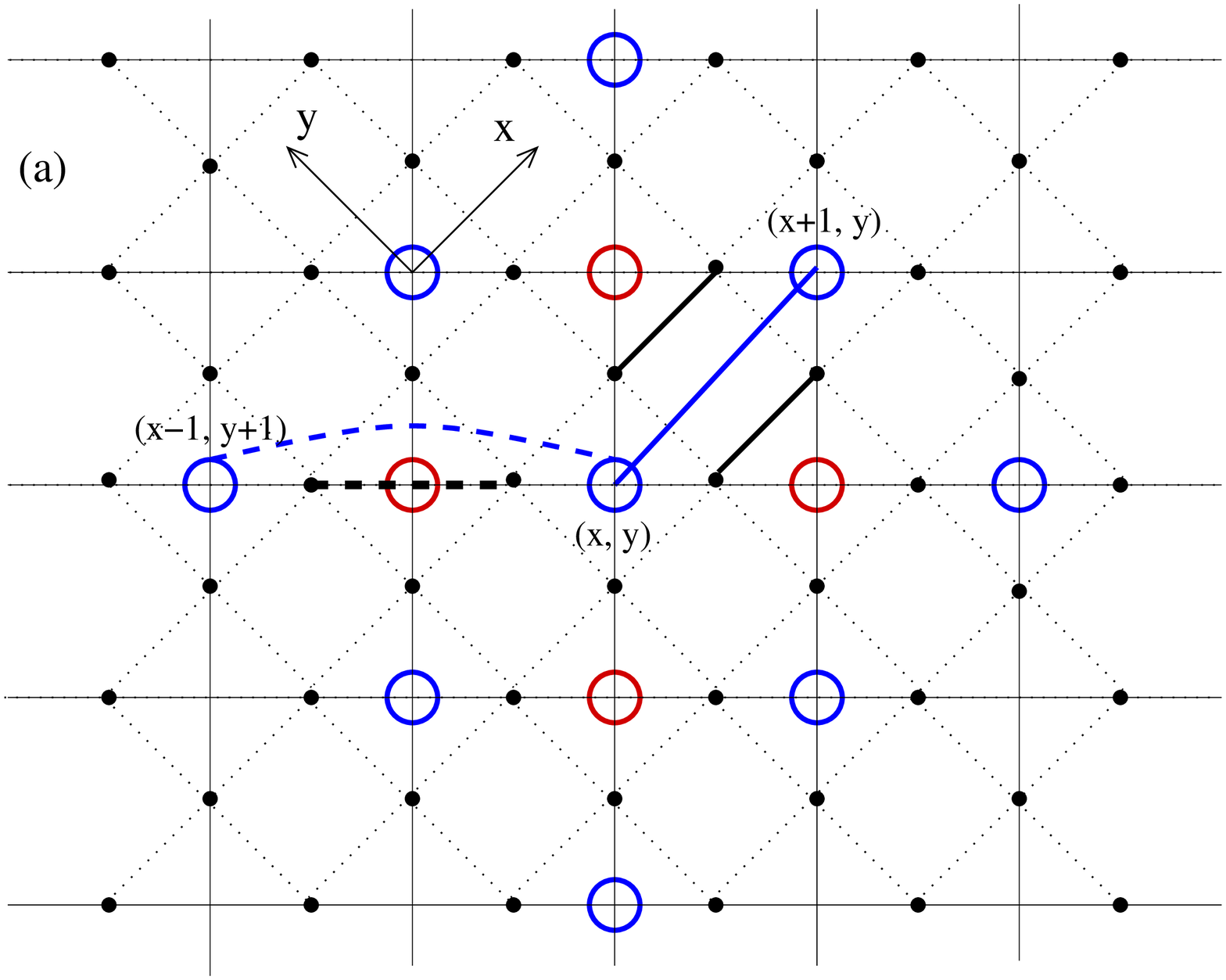}
    \includegraphics[width=0.9 \linewidth]{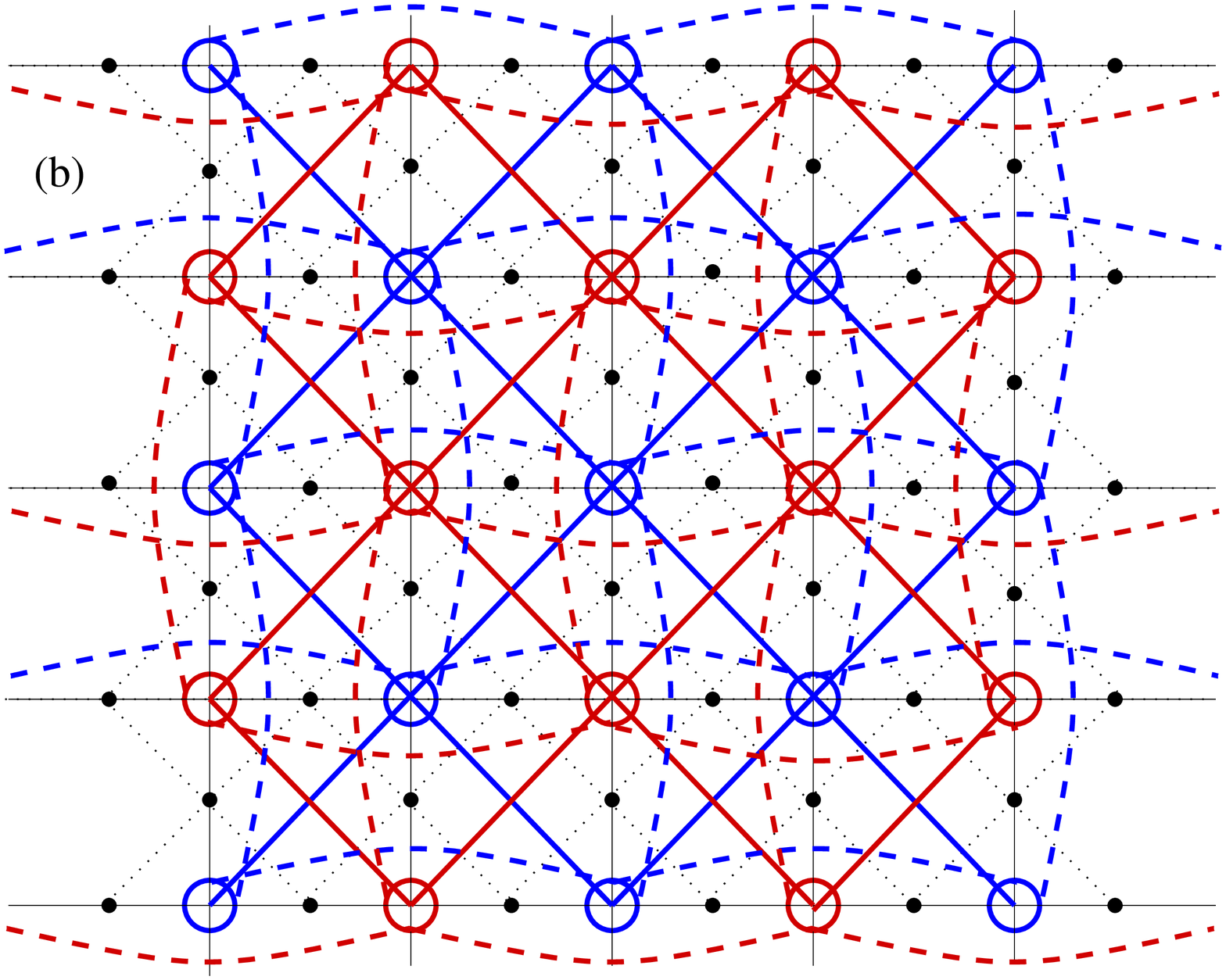}
    \caption{(color online) (a) Diagonal-Ising interactions (solid-black lines in the upper right part) and
      its corresponding effective interaction on quasi-spins (solid-blue line). Horizontal-Ising interaction
      (dashed-black line in the left part) and its corresponding effective interaction on quasi-spins (dashed-blue line).
      Blue and red circles represent the quasi-spins. (b) Two sublattices (blue and red), which interact like
      TFI model with next-nearest neighbor interactions as an effective theory to describe the toric code Hamiltonian
      in the presence of full-Ising interactions.}
    \label{Effective-2D}
  \end{center}
\end{figure}

The diagonal-Ising interaction does not change the state of a vertex that shares two spins while
it flips the state of two vertices (along diagonal direction) that share only one spin.
The right upper part of Fig.~\ref{Effective-2D}-(a) shows two quasi-spins denoted by blue circles, which
are being flipped according to the diagonal-Ising terms (shown by diagonal-solid black lines). The effective interaction
is represented by the solid-blue line in the diagonal direction. Moreover, the four vertices around a rhombus
are decomposed to two decoupled sublattices shown by blue and red circles. Hence, the effective
Hamiltonian for the toric code in the presence of diagonal-Ising interactions is given by
\be
{\cal H}_{eff}^{TD}=-2 J_z\sum_{<\nu, \nu'>} T^x_{\nu} T^x_{\nu'} - J_v \sum_{\nu} T^z_{\nu},
\label{HeffTD}
\ee
where $<\nu, \nu'>$ runs on the nearest neighbor quasi-spins of either blue or red sublattices.
The lattice of quasi-spins (either blue or red) is a two-dimensional square lattice rotated $\pi/4$
with respect to the original lattice, namely shown by the x-y unit vectors in Fig.~\ref{Effective-2D}-(a).
This is in complete agreement with the effective model presented in Ref.\onlinecite{Karimipour:2013},
which shows a quantum phase transition from the $\mathbb{Z}_2$ spin-liquid ground state of topological toric code ($J_z=0$)
to a ferromagnetic product state of Ising limit ($J_v=0$).
According to the Monte-Carlo~\cite{Blote:2002} and DMRG~\cite{PhysRevB.57.8494} simulations
on 2D TFI model,
the quantum critical point is at $J_v=6.09 J_z$.
Similar behavior is expected for the antiferromagnetic diagonal-Ising interactions,
which can be obtained by the transformation $J_z \rightarrow -J_z$ and $\sigma^z_i \rightarrow -\sigma^z_i$
for $i \in \Sigma_A$, where $\Sigma_A$ is one of the sublattices of the original bipartite two-dimensional lattice.
Needless to mention that the AF Ising limit presents a N\'{e}el ordered state.

\subsection{Horizontal-Vertical Ising interactions}

A bond of horizontal (or vertical) Ising interaction does not change the state of a
quasi-spin, which shares two spin at the corresponding vertex. This is shown by the tick-dashed-black
line in the left part of Fig.~\ref{Effective-2D}-(a), which crosses a red circle. The quasi-spin
denoted by the red circle is not changed, while the quasi-spins of the horizontal neighboring blue circles are being
flipped. Hence, a horizontal (vertical) Ising interaction  flips the nearest neighbor quasi-spin
of a sublattice oriented horizontally (vertically). The lattice of quasi-spins denoted by
blue and red circles in Fig.~\ref{Effective-2D}-(a) is bipartite, where the blue and red circles
represent the two sublattices.
Thus, the effective model of toric code
with HV-Ising interactions is given by the following Hamiltonian
\be
{\cal H}_{eff}^{THV}= -J_z\sum_{<\nu, \nu'>} T^x_{\nu} T^x_{\nu'} - J_v \sum_{\nu} T^z_{\nu},
\label{HeffTD}
\ee
in which $<\nu, \nu'>$ defines nearest neighbor quasi-spins of either blue or red sublattices.
Each sublattice forms a two-dimensional square lattice where the lattice constant is twice as the original one.
According to Ref.\onlinecite{Blote:2002}, the QPT takes place at $J_v=3.044 J_z$
from the $\mathbb{Z}_2$ spin-liquid state ($J_z=0$) to the ferromagnetic product state.
Similar to the previous case, the phase diagram is symmetric with respect to $J_z \rightarrow -J_z$,
which covers the antiferromagnetic regime. A remark is in order here, the original (and also the effective lattice)
is bipartite for both D-Ising and HV-Ising interactions at
the extreme limit of Ising interactions ($J_v=0$). This avoids frustration for the antiferromagnetic Ising couplings,
which is not preserved for the full-Ising interactions as will be discussed next.

\subsection{Full-Ising interactions}

The effective theory of toric code in addition to full-Ising interactions can be obtained
by considering the representations of both diagonal and HV-Ising interactions in the quasi-spin subspace.
The following facts help to realize easily the structure of the  effective theory. (i) The quasi-spin lattice
(of blue and red circles) is bipartite such that neither a diagonal nor an HV-Ising interaction
make an interaction between the blue and red sublattices. (ii) The quasi-spin representation forms
a square lattice whose lattice spacing is $\sqrt{2}$ times larger than the original one and is rotated $\pi/4$
with respect to the original lattice. The unit-vectors of the quasi-spin lattice are labeled by
x-y in Fig.~\ref{Effective-2D}-(a). (iii) The diagonal-Ising interactions establish the nearest neighbor
interactions of strength $2J_z$ on each sublattice. (iv) The HV-Ising interactions build up the next-nearest
neighbor (diagonal) interactions of strength $J_z$ on each sublattice. Accordingly, the
effective model of the toric code in the presence of full-Ising interactions is
\bea
{\cal H}_{eff}^{TFI}=&&-2 J_z\sum_{<\nu, \nu'>} T^x_{\nu} T^x_{\nu'} - J_v \sum_{\nu} T^z_{\nu} \nonumber \\
&&-J_z\sum_{<<\nu, \nu'>>} T^x_{\nu} T^x_{\nu'} ,
\label{Heff-ful}
\eea
where $<\nu, \nu'>$ stands for NN and $<<\nu, \nu'>>$ for NNN quasi-spins.
The effective Hamiltonian is a two-dimensional NNN TFI model on
a square lattice that is shown in Fig.~\ref{Effective-2D}-(b), where the blue and red colors show two decoupled
sublattices. The solid lines show NN interactions while the dashed one show
NNN ones.

In the ferromagnetic regime ($J_z>0$) a quantum phase transition occurs from the topological toric code
ground state ($J_z=0$) to the symmetry broken ferromagnetic state. The corresponding QCP 
is at $J_v=12 J_z$ within the mean-field approximation.

The antiferromagnetic regime, $J_z<0$, would be essentially different from the ferromagnetic one
as a result of frustration in the AF Ising limit ($J_v=0$). The frustration arises from the
antiferromagnetic bonds, which form triangles in the original lattice. The ground-state manifold
is highly degenerate, which is composed of triangles having 2-up-1-down or 1-up-2-down spins (in
z-direction). This is a manifestation of {\it classical spin-liquid} phase on a two-dimensional
lattice. It suggest a quantum phase transition from the classical spin-liquid phase ($J_v=0$) to the
$\mathbb{Z}_2$ topological spin-liquid phase ($J_z=0$) at a finite ratio of $J_z/J_v$.
However, according to the effective theory presented in Eq.\ref{Heff-ful}, we expect
an order-by-disorder phase transition before the transition to the topological spin-liquid phase.
At the AF Ising limit ($J_v=0$)
the effective theory is known as $J_1-J_2$ Ising model on square lattice with $J_2=0.5 J_1$,
where $J_1$ is NN and $J_2$ is NNN antiferromagnetic interactions.
It is known that $J_2=0.5 J_1$ is the critical point of $J_1-J_2$ Ising model, which separates the
N\'{e}el ordered phase ($J_2<0.5 J_1$) from the collinear ordered one ($J_2>0.5 J_1$).
The critical nature of $J_2=0.5 J_1$ of our effective theory manifests the frustration induced by triangles
of the original model, which leads to an extensive degeneracy in both the effective ground state and
the original one. The onset of transverse magnetic field ($\Gamma$) at the critical point $J_2=0.5 J_1$ of
the effective theory leads to an order-by-disorder~\cite{Sadrzadeh:1405.1233}, which looks like a
ferromagnetic order for small fields, $\Gamma \lesssim 0.2 J_1$,
before a transition to the fully polarized state.
This is in agreement with the order-by-disorder transition proposed for the
fully-frustrated Ising model on the square lattice~\cite{PhysRevLett.84.4457}
A ferromagnetic state of the effective theory,
$\langle T^x_{\nu} \rangle =1$, corresponds to $A_v=1$, in which only those spin configurations
of two-up-two-down on a vertex contribute to the ground state, which can be called
a {\it resonating ice-state} resembling the ice-rule configuration of spin-ice.
The spin-ice configurations do not include the all-up or all-down states of
$A_v=1$ subspace. Hence,
the classical spin-liquid state is unstable against quantum fluctuations at $J_v=0$ toward
a resonating ice-state for
$0 < J_v \lesssim 0.2 J_z$. Increasing the vertex coupling makes a transition to the
topological $\mathbb{Z}_2$ spin-liquid state for $J_v \gtrsim 0.2 J_z$.
More investigations is required, which needs demanding resources for
numerical computations that is beyond the scope of present manuscript.

\section{Summary and discussion \label{summary}}
\label{Summary}

We have studied the Kitaev Hamiltonian (${\cal H_K}$) on a ladder geometry. We find that the
ground state of Kitaev ladder is an SPT phase protected by $\mathbb{Z}_2 \times \mathbb{Z}_2$ symmetries namely
${\cal X}=\prod_i \sigma_i^x$ which runs over all ladder bonds and
${\cal Z}=\prod_{\ell \notin rungs} \sigma_{\ell}^z$
that excludes the rung bonds. We have justified our argument by employing iDMRG method within an
iMPS representation, which leads to
inequivalent projective representation of $\mathbb{Z}_2 \times \mathbb{Z}_2$ symmetries providing
the phase factor order parameter ${\cal O}=-1$, for the Kitaev phase.

We have also investigated the competition between the Kitaev and Ising terms on the ladder,
which is given by deriving the corresponding effective theory in addition to the direct iDMRG computations.
For the Ising interactions being solely on the edges of rhombus or on the legs, the effective Hamiltonian
is given by decoupled one-dimensional TFI models, which explains the quantum phase transition from the
Kitaev SPT phase to the antiferro/ferro-magnetic phase at the exact finite value $J_z/J_v=0.5, 1.0$, respectively.
The quantum phase transition is justified by numerical divergence of entanglement entropy ($S_E$)
at the critical point, the change in the degeneracy of the entanglement spectrum, ground-state fidelity,
and magnetic order parameters. The quantum critical points and their corresponding central charges
of the effective theory and iDMRG results agree with each other, exactly. The critical behavior
of Kitaev-rhombic-Ising interactions is given by central charge $c=1$, while the Kitaev-leg-Ising ladder
is represented by $c=2$. If the Ising interactions reside on both rhombus and legs of ladder, the effective
theory would be NNN TFI chain with the critical properties given by $c=1$.
For the ferromagnetic Ising interactions it leads to a quantum
phase transition at finite ratio $J_z/J_v=0.28$, while for the
antiferromagnetic Ising interactions our data shows a broad range of finite entanglement scaling.
It is the interplay between the classical spin-liquid at AF Ising limit
and the Kitaev SPT phases. According to the effective theory, the onset of Kitaev
term induces quantum fluctuations in the classical spin-liquid subspace, which would finally lead to the
Kitaev SPT phase passing through an intermediate floating phase. The Kitaev SPT phase is
persistent for $|J_z|/J_v \lesssim 3$ witnessed by the even degeneracy of the entanglement spectrum.

A remark is in order concerning the effective theory introduced in this paper.
For simplicity, we consider the ladder geometry with periodic boundary condition along the legs,
where $N$ is the number of spins sitting on each leg or rung of ladder, which sums up to $3N$ spins.
The dimension of Hilbert space of the original ladder is $2^{3N}$. The number of triangles
(quasi-spins) is $2N$ and the number of rhombuses (plaquettes) is $N$. Accordingly, the
dimension of the Hilbert space of the effective theory (in terms of quasi-spins) is $2^{2N}$
which is smaller than the original Hilbert space by a factor of $2^{N}$.
In principle, this is always the case for an effective theory, which is responsible
for the low-energy behavior of the original model
and is confirmed by numerical iDMRG results.
However, taking into account the plaquette degrees of freedom  ($B_p=\pm 1$)
we find the lost $2^{N}$ degrees of freedom. For the lowest energy
spectrum we consider all $N$-plaquettes to be at $B_p= +1$, which adds a constant term $-J_p N$ to
the effective theory. As far as all configurations of the original spin model have been kept in constructing
the effective theory we expect that the whole spectrum of the original model is represented by
a tower of TFI models in addition to their corresponding constant values, i.e., $(- J_p) \sum_p B_p$.

\begin{figure}[tb!]
  \begin{center}
    \includegraphics[width=0.99 \linewidth]{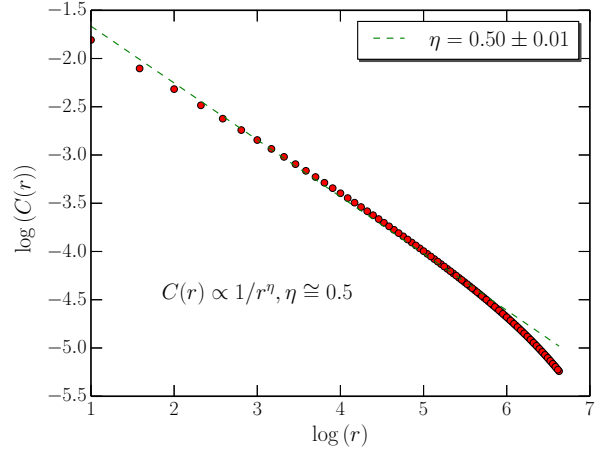}
    \caption{(color online) Log-log plot of correlation function $C(r)$ versus $r$
for the Kitaev-rhombic Ising ladder at the QCP. The (green) dotted line shows
the best fit of $r^{(-0.5)}$, which states $\eta= 0.50\pm0.01$.
The correlation length is $\xi\backsimeq400$ for $\chi=150$, which determines the
reliable behavior for $r<\xi$, i.e. $\log(r)\lesssim 5.99$.
}
    \label{correlation}
  \end{center}
\end{figure}

Having in mind that the effective theory considers the whole degrees of freedom of the original model,
we find the exact critical exponents of the mentioned QCPs. For instance, we calculate in detail the
magnetization exponent ($\beta$) close to transition of the ferromagnetic phase of Kitaev RI-ladder
(Sec.~\ref{RI-section}). The ferromagnetic order parameter $\langle \sigma_z \rangle$ is
\be
\langle \sigma_z \rangle=\frac{1}{3N}\sum_{i=1}^N \langle \psi_0 |
\sigma_1^z(i)+\sigma_3^z(i)+\sigma_3^z(i) | \psi_0 \rangle.
\label{m_z}
\ee
The effect of $\sigma_1^z(i)$ on the ground state of ladder ($| \psi_0 \rangle$) is
equivalent to flip the state of quasi-spins denoted by the two triangles, which share
$\sigma_1^z(i)$ at their common corner. It is shown by
\be
\sigma_1^z(i) | \psi_0 \rangle = \tau_{\mu}^x | \varphi_0^{(even)} \rangle
\otimes \tau_{\mu-1}^x | \varphi_0^{(odd)} \rangle,
\label{sz-tax2}
\ee
where $| \varphi_0^{(even)} \rangle$ ($| \varphi_0^{(odd)} \rangle$) represents the ground state
of TFI effective theory for even (odd) decoupled chain. Hence, we conclude that
\be
\langle \sigma_z \rangle=\big(\langle \tau^x \rangle_{TFI}\big)^2= |J_z-J_z^c|^{\frac{1}{4}},
\label{mag-exponent}
\ee
which leads to $\beta=1/4$ as confirmed numerically in the inset of Fig.\ref{S_R}.
A similar calculation gives the exponent of the algebraic decay of correlation
functions at the QCP,
\be
C(r)|_{J_z=J_z^c}=\langle \sigma_1^z(i) \sigma_1^z(i+r) \rangle \sim \frac{1}{r^{\eta}}, \hspace{5mm} \eta=\frac{1}{2}.
\label{cf}
\ee
This is in agreement with the numerical computation of correlation function of the
Kitaev RI-ladder performed at $J_z=J_z^c=0.5$ in Fig.~\ref{correlation}, where
the numerical exponent is $\eta = 0.50\pm0.01$.
It should be noticed that both $\beta$ and $\eta$ exponents are twice as the corresponding
one of the TFI chain. It means that the decoupled chains of the effective theory contribute
to the quantum critical properties of the original ladder. In other words, although the
effective TFI chains are decoupled they are not independent.
Our calculations for all critical points of AF/F Kitaev RI, LI and F RLI cases give
the same exponents, namely $\beta=1/4$ and $\eta=1/2$.
The case of AF RLI needs more delicate considerations being close to a Kosterlitz-Thouless transition,
which needs more extensive numerical computations out of the scope of this article.

The extension of our approach to the two dimensional case would lead to study the competition
between the topological $\mathbb{Z}_2$ spin-liquid state of the toric code with the symmetry broken
or classical spin-liquid state of Ising interactions.
If the Ising interactions exist either on the diagonal or horizontal-vertical direction of
the two-dimensional lattice, the QPT between topological ground state of toric code and
the ferro/antiferro-magnetic phase is given by the TFI model on two-dimensional square lattice.
However, in the case of both diagonal and horizontal-vertical Ising interactions the effective
theory is NNN TFI on 2D square lattice. For the ferromagnetic Ising interactions, it leads to
a quantum phase transition at finite ratio $J_v/J_z$, while for the antiferromagnetic case 
we expect an order-by-disorder transition at $J_v=0$ followed by a transition to the 
spin-liquid state at $J_v/J_z \simeq 0.2$.
The frustration induced by the antiferromagnetic coupling
hinders the long-range order of the Ising limit toward an exponentially degenerate ground state
configurations called classical spin-liquid. According to the effective theory, the onset of
Kitaev toric code term perturb the model toward an ordered phase for $J_v/J_z \lesssim 0.2$
and finally the $\mathbb{Z}_2$ spin-liquid state appears for  $J_v/J_z \gtrsim 0.2$.

\begin{acknowledgments}
The authors would like to thanks
S. Bhattacharjee, J. Kj\"{a}ll,
S. Moghimi-Araghi, O. Petrova, F. Pollmann and A. T. Rezakhani
for fruitful discussions and comments.
  This work was supported in part by
  the Office of Vice-President for Research of
  Sharif University of Technology.
  A. L. gratefully acknowledges the Alexander von Humboldt Foundation for financial support.
\end{acknowledgments}

\appendix
\section{Numerical approach: iDMRG}
\label{iDMRG}
To examine the properties of the model, we have made use of the
standard iDRMG technique that is based on an infinite matrix product
state (iMPS) representation for the ground state\cite{Mcculloch:2008,
  Schollwock:2011}. It is a well known fact that iMPS is an efficient
method to describe
translationally invariant many body states with an
accuracy depending on the dimension of implemented matrices \cite{Verstraete:2006,Hastings:2004}.

The translationally invariant ground state is characterized by
canonical \cite{PhysRevB.78.155117} $\Gamma$ and $\Lambda$ matrices,
\begin{equation}
|\Psi \rangle=\sum\limits_{
s_1,\ldots,s_N
} \mathrm{Tr} [\Gamma^{s_{1}}\Lambda \dots \Gamma^{s_{N}}\Lambda]|s_{1}\dots s_{N}\rangle,
\label{State}
\end{equation}
which
satisfy the following (fixed point) relation,
\be
 \sum_s \Gamma^s \Lambda^s\Lambda^{s \dagger}\Gamma^{s \dagger} =\openone,
\ee
where the sum is over different spin configurations and $\Gamma$s
serve as the matrix coefficients for these configurations.

The spectrum (i.e.\ singular values in the Schmidt
decomposition of the left and right bipartition of the Hilbert space)
is simply the square root of $\Lambda$ and the entanglement entropy
is defined as,
\be
S_{\text{E}} =-\sum_i \Lambda_{ii}^2 \log{(\Lambda_{ii}^2)}.
\ee

In the general case, to give an exact representation of a state,
iDMRG needs infinitely large matrices. Hopefully, this is not
necessary, specially in the case of gapped systems, by putting an
upper bound on the cardinality of the matrices $\chi$, and truncating
the spectrum, one can reach a good approximation, which has all the properties of
the low energy state. This will give rise to the
so-called, truncation error which can be controlled by the dimension
of matrices and is the cause of the entropy scaling.

Since the ground state is known to be gapped for the extreme coupling limit of our model,
it can be represented by a finite iMPS, where the truncation error for
the couplings that are far from the critical point, is less than
machine precision. However, close to the critical point, when the
ground state entanglement spectrum should show a long tail, the truncation errors
become considerable and they do not vanish even when we increase the
size of matrices.

After reaching the canonical $\Lambda$, $\Gamma$ with the desired
accuracy, several properties of the ground state can be evaluated using
the iMPS representation. It includes any local observable
like energy and $\sigma^z$, entanglement spectrum and
the corresponding von-Neumann entropy, the application of symmetry
operators, and the ground state fidelity\cite{PhysRevLett.100.080601}. In
order to calculate the mentioned quantities the concept of transfer matrix
should be introduced,
\be
T_{\alpha\alpha',\beta\beta'} = \sum_{s} (\Gamma_{\alpha\beta}^s
\Lambda_{\beta}) (\Gamma_{\alpha'\beta'}^s  \Lambda_{\beta'})^{\ast}.
\ee
Expectation value of a local operator (defined on one specific site), such as $\hat{{\cal O}}$, is obtained by using the following
transfer matrix,
\be
\widehat{T}_{\alpha\alpha',\beta\beta'} = \sum_{s, s'} (\Gamma_{\alpha\beta}^s
\Lambda_{\beta}) \hat{{\cal O}}^{s,s'}(\Gamma_{\alpha'\beta'}^{s'}  \Lambda_{\beta'})^{\ast}.
\ee
Expectation value of $\langle\Psi| \hat{{\cal O}} |\Psi\rangle$ simply reduces to $\mathrm{Tr}(\Lambda\otimes \Lambda  \widehat{T})$.
For the fidelity calculations, iDMRG should calculate the canonical iMPS for two
very close couplings ($J_z^-, J_z^+$),
\bea
&J_z^- = J_z - \delta/2, \nonumber \\
&J_z^+ = J_z + \delta/2
\eea
For an iMPS state the fidelity is defined as the largest
eigenvalue of the transfer matrix constructed as the product of two
close couplings,
\be
\overline{T}_{\alpha\alpha',\beta\beta'}^{\delta} = \sum_s (\Gamma_{\alpha\beta}^s
\Lambda_{\beta})_{J_z^-} (\Gamma_{\alpha'\beta'}^s
\Lambda_{\beta'})^{\ast}_{J_z^+}.
\ee

The central charge is calculated according to the scaling relation
between von-Neumann entropy and the correlation length. The correlation
length is defined as the second largest eigenvalue ($e_{2}$) of the transfer matrix $T$ ~\cite{Kjall:2013,Calabrese:2004},
\be
S_{\text{E}} \propto \frac{c}{6} \log{(\xi)}, \qquad \xi =-\frac{1}{\ln{(e_2)}}.
\ee

The general scheme of the algorithm for our model is as
follows. First, we formulate the model to get a 1D model with only NN interactions, we
bundle every three particles on a triangle, as shown in the Fig.~\ref{Kitaev-ladder},
into one unit-cell with dimension 8.
For instance,
according to the definition in Eq.\ref{Sigma}
the Kitaev Hamiltonian is written in the following form
\be
{\cal H_K} = -J_v \sum_{<i,j>}\Sigma(i)^{IxI}\Sigma(j)^{xxI}
-J_p \sum_{<i,j>} \Sigma(i)^{zzz}\Sigma(j)^{zII}.
\ee
Similar expressions would be used for the Ising terms.
The CPU time of iDMRG algorithm is proportional to the square of
the spin dimensions $d^2$. It is obvious that the calculation time
needed to accomplish a simple iDMRG on the mentioned lattice (with $d=8$) is much larger
than iDMRG performed on a lattice with spin dimension 2. This is
the main reason that we were unable to examine larger matrices for
this model. For example, the necessary time for convergence was about couple of weeks
for a single run of matrix size $\chi=128$, and close to critical region.
The convergence criterion was a fixed point
relation between the $\Lambda$ generated at the current step with the
$\Lambda$ of the last step.

We have also examined the iDMRG for one-dimensional NNN TFI model and compared the results
with the corresponding one of the original ladder.
The entanglement spectrum and hence the entropy
was the same within relative error of $10^{-5}$.

\section{Symmetry}
\label{symmetry}
A symmetry is defined as an operation, which leaves the model Hamiltonian invariant.
These symmetries can form either an ordinary group or a projective one.
However, if the ground state of Hamiltonian does not respect
the Hamiltonian symmetries, one concludes the phase is a symmetry
broken one. At the same time, the remaining symmetry groups
can protect a phase due to their inequivalent projective representations,
also known as symmetry fractionalization. These two
properties can be used to assign a unique label to every possible
phase of a system and to detect possible phase transition within this
classification \cite{XieChen:2011}.

The ground state of the Kitaev ladder Hamiltonian (${\cal H_K}$) is
doubly degenerate and both ground states are invariant under the operations
of ${\cal X}$ and ${\cal Z}$ (defined in Eq.\ref{XZ}). The mutual symmetry operation ${\cal X} \times {\cal Z}$ defines a $\mathbb{Z}_2\times \mathbb{Z}_2$ symmetry, which protects the Kitaev ladder ground states. In other words, providing
the symmetry is preserved, the Kitaev SPT phase can not be adiabatically mapped to a fully product state.

To gain more insight on how the $\mathbb{Z}_2\times \mathbb{Z}_2$ symmetry group can serve to protect the
degeneracy of the entanglement spectrum, and  how we can express them numerically we need to
explore the properties of the symmetry group in terms of iMPS representation.
To preserve the $\mathbb{Z}_2\times \mathbb{Z}_2$ symmetry for an iMPS state,
the following relation should be satisfied ~\citep{PhysRevA.79.042308},
\be
 \sum_{s'} u_{ss'}(g) \Gamma^{s'}_{\alpha\alpha'} =
U^{\dagger}(g)_{\alpha\beta} \Gamma^{s}_{\beta\beta'} U(g)_{\beta'\alpha'},
\label{TransformationMPS}
\ee
where $u(g)\in G$, $G=\{\Sigma^{Izz}, \Sigma^{xxx}, -\Sigma^{xyy}, \Sigma^{III}\}$
and $g$ represents the index of group elements. To obtain $U_g$ for all elements of
the group $G$, we construct the following transfer matrix ($\widehat{T}^{g}$) for each element of group $G$,
\be
\widehat{T}^{g}_{\alpha\alpha',\beta\beta'} = \sum_{s, s'} (\Gamma_{\alpha\beta}^s
\Lambda_{\beta}) u(g)^{s,s'} (\Gamma_{\alpha'\beta'}^{s'}  \Lambda_{\beta'})^{\ast}.
\ee
The symmetry represented by $u(g)$ on all sites is
respected, if the
largest eigenvalue of $\widehat{T}^{g}$ becomes equal to 1. Using Eq.~\ref{TransformationMPS},
one can show the corresponding
eigenvector is simply $U^{\dagger}_g$~\citep{Pollmann:2012}.

Generally, $U_g, U_{g'}$ may not always form a
regular group but a projective one. To see this behavior we need to
apply the symmetries in different order and make use of the facts that
$u(g)u(g') = u(g')u(g)$ and $u^{2}(g)=\openone$. Using $u(g)u(g') = u(g')u(g)$, we conclude
\bea
& u(g)u(g') \Gamma = U_g U_{g'} \Gamma U^{\dagger}_{g'}U^{\dagger}_g, \nonumber \\
& u(g')u(g) \Gamma = U_{g'} U_g \Gamma U^{\dagger}_{g}U^{\dagger}_{g'},\nonumber \\
& \Rightarrow U_gU_{g'} = e^{i\Omega_{gg'}} U_{g'}U_g, \label{Projective}
\eea
where the phase $e^{i\Omega_{gg'}}$ is called ``phase factor" (for simplicity we drop indices corresponding to summations).
The property of $u^{2}(g)=\openone$ results in $U^{2}(g)=e^{i\theta_{g}}\openone$.
Using Eq.~\ref{Projective} and $U^{2}(g)=e^{i\theta_{g}}\openone$, one can easily show
that $e^{i\Omega_{gg'}}$ can only be $\pm 1$. The signs introduce two different kind of orders,
i.e. SPT and trivial orders. Throughout the SPT (trivial) phase, $e^{i\Omega_{gg'}}=-1(+1)$ and
only upon quantum phase transition, the sign can change. The two signs also represent two inequivalent
projective representations of $\mathbb{Z}_2\times \mathbb{Z}_2$ symmetry.

One can exploit this property and define an order parameter
${\cal O}$ (called phase factor order parameter), which can serve to detect, by measuring the sign, which projective
representation holds for a possible phase,

\be
{\cal O} = \frac{1}{\chi} \text{Tr}\left(U_gU_{g'}U^{\dagger}_gU^{\dagger}_{g'}\right).
\ee

When the iMPS doesn't possess one of the symmetries in the group, the
phase factor order parameter ${\cal O}$ is simply $0$, demonstrating symmetry broken phase.

The proposed $\mathbb{Z}_2\times \mathbb{Z}_2$ symmetry group is not just the symmetry
for the Kitaev Hamiltonian, but it commutes with the Ising
interactions as well. As a result the phase
factor order parameter ${\cal O}$,
can be a good quantity to  observe the phase transition which
kills the symmetry protected phase.  When the system is close to the
Kitaev phase the phase factor order parameter has a negative sign, which shows the system is in a symmetry protected
state, but for the ferromagnetic phase, while
$\Sigma^{Izz}$ is still respected,  $\Sigma^{xxx}$ is no longer
preserved and the phase operator order parameter suddenly drops to zero, as
shown in Fig.~\ref{phases}.







\bibliography{kitaev-lattice}

\begin{thebibliography}{61}
\bibitem [{Pro()}]{Projective-note}%
  \BibitemOpen
  \href@noop {} {}\bibinfo {note} {Projective representation of a specific
  symmetry group, e.g., $G$ is accompanied by a phase factor. For instance, if
  elements $g, g'$ and $g''\in G$ have the relation $gg'=g''$, the
  corresponding projective representation becomes
  $U_{g}U_{g'}=e^{i\theta}U_{g''}$.}\BibitemShut {Stop}%
\end{thebibliography}

\end{document}